\documentclass{article}

\usepackage{arxiv}

\usepackage[T1]{fontenc}
\usepackage{url}
\usepackage{booktabs}
\usepackage{array}
\usepackage{amsfonts}
\usepackage{amsmath}
\usepackage{amsthm}
\usepackage{nicefrac}
\usepackage[expansion=false]{microtype}
\usepackage{graphicx}
\usepackage[numbers,sort&compress]{natbib}
\usepackage[scr=rsfso,frak=euler]{mathalfa}
\usepackage{xargs}
\usepackage[dvipsnames]{xcolor}
\usepackage{enumitem}
\usepackage{algorithm}
\usepackage{algpseudocode}

\usepackage{hyperref}
\usepackage{orcidlink}
\usepackage{doi}
\graphicspath{{figures/}{./}}

\usepackage{tikz}
\usetikzlibrary{positioning,arrows.meta,calc}
\definecolor{ink}{HTML}{2B2B2B}\definecolor{grey}{HTML}{8A8A8A}
\definecolor{es}{HTML}{C8553D}\definecolor{ent}{HTML}{3F72AF}\definecolor{law}{HTML}{4F8A6B}
\usepackage{etoolbox}
\makeatletter
\newif\if@firstsectiondone
\newif\if@samepage
\newcommand{\samepagesection}{\global\@samepagetrue}
\pretocmd{\section}{%
  \if@samepage\global\@samepagefalse\else
  \if@firstsectiondone\clearpage\else\global\@firstsectiondonetrue\fi\fi}{}{%
  \PackageError{main}{could not patch \string\section}{}}
\pretocmd{\subsection}{\if@nobreak\else\clearpage\fi}{}{%
  \PackageError{main}{could not patch \string\subsection}{}}
\makeatother
\usepackage{tabularx}
\newcolumntype{L}{>{\raggedright\arraybackslash}X}
\newcolumntype{P}[1]{>{\raggedright\arraybackslash}p{#1}}
\AtBeginEnvironment{table}{\small\setlength{\tabcolsep}{5pt}}

\newtheorem{problem}{Problem}

\newtheorem{proposition}{Proposition}
\theoremstyle{remark}

\theoremstyle{plain}

\DeclareMathOperator*{\argmax}{arg\,max}

\title{Distributional Deep Hedging}
\date{}

\author{%
  El-Amine Cherrat\,\orcidlink{0009-0006-8682-3287} \\
  Quantum Signals SAS \\
  Paris, France \\
  \texttt{amine@quantumsignals.ai} \\
  \And
  Skander Kazdaghli\,\orcidlink{0009-0005-9044-0919} \\
  Quantum Signals SAS \\
  Paris, France \\
  \texttt{skander@quantumsignals.ai} \\
  \And
  Iordanis Kerenidis\,\orcidlink{0000-0003-0659-3727} \\
  IRIF, CNRS, Universit\'e Paris Cit\'e, France \\
  Quantum Signals SAS, Paris, France \\
  \texttt{iordanis@quantumsignals.ai} \\
}
\renewcommand{\shorttitle}{Distributional Deep Hedging}

\definecolor{linkblue}{HTML}{1D3D6E}
\hypersetup{
colorlinks=true, linkcolor=linkblue, citecolor=linkblue, urlcolor=linkblue,
pdftitle={Distributional Deep Hedging},
pdfauthor={El-Amine~Cherrat, Skander~Kazdaghli, Iordanis~Kerenidis},
}

\begin{document}
\maketitle

\begin{abstract}
Deep hedging learns a hedging strategy by optimising a risk preference over a neural-network policy. The
preference is fixed before training, so each new preference requires a new policy and, in actor--critic
methods, a new critic. We propose distributional deep hedging, in which the distribution of the hedging
profit and loss is learned through its quantiles and the risk preference is applied afterwards, as a
weight on those quantiles. We characterise the preferences that admit this representation, which include
expected shortfall and its mixtures, and show that the entropic utility is computed from the same
quantiles. At the current policy, each of these preferences is bounded below by an expected utility with
the same gradient. This bound reduces risk-sensitive hedging to a sequence of risk-neutral problems, from
which we derive policy-search and actor--critic algorithms. In numerical experiments, the learned hedges
approach references computed by dynamic programming, and a single distributional critic trained without
a preference can serve preferences chosen after training.
\end{abstract}

\section{Introduction}
\label{sec:introduction}

Hedging a derivative in a market with trading frictions has no closed-form solution in general. Exact
replication survives in particular models, such as the binomial model with proportional costs
\citep{boyle1992-option-replication-discrete-time} or models with price impact and execution costs
\citep{itkin2026-when-hedging-changes-payoff}. Between two trading dates, however, a price usually has more
possible moves than there are instruments to trade, so a claim cannot in general be replicated, and the
hedger has to weigh the cost of trading against the risk left in the book. Deep hedging
\citep{buehler2019-deep-hedging} addresses this by representing the hedging strategy as a neural network
and optimising a risk preference of the terminal profit and loss directly over simulated paths. The
approach does not depend on the model that generates the paths, and it has since been extended to
actor--critic methods \citep{murray2022-deep-hedging-continuous-rl,buehler2022-deep-bellman-hedging}, in
which a critic estimates the value of the preference along the way.

In all of these methods the risk preference is fixed before training. The policy is optimal for that
preference only, and the critic of an actor--critic method estimates a value defined relative to it, so
changing the preference means training again. A trading desk, however, rarely works with a single
preference. It reports expected shortfall at prescribed levels, sets internal limits that weight several
depths of the tail, and revises its risk aversion over time (Section~\ref{subsec:ddh}); each of these
evaluations requires its own trained object, and none of them reuses what was learned for another.

\begin{figure}[t]
\centering
\resizebox{0.8\textwidth}{!}{%
\begin{tikzpicture}[font=\sffamily\small, text=ink,
  card/.style={rounded corners=4pt, draw=#1, line width=0.8pt, fill=#1!8, minimum height=30mm, inner sep=0pt},
  arr/.style={-{Latex[length=2.6mm,width=2mm]}, line width=1pt, draw=grey}]
\node[card=grey, minimum width=27mm] (m) at (0,0) {};
\node[anchor=north] at ($(m.north)+(0,-1.6mm)$) {\bfseries Market};
\begin{scope}[x=1mm,y=1mm,shift={($(m.center)+(-10mm,-6mm)$)}]
  \foreach \s/\c in {0/0.15,1/0.35,2/-0.1,3/0.5,4/0.05}{
    \draw[grey!70, line width=0.5pt] plot[smooth, domain=0:20, samples=12] (\x, {6+3*sin(\x*25+\s*70)*\c+ \s*0.6+ 2*sin(\x*9+\s*40)});}
\end{scope}
\node[anchor=south, font=\sffamily\scriptsize, text=grey] at ($(m.south)+(0,1.3mm)$) {simulated paths};
\node[card=law, minimum width=42mm, right=12mm of m] (l) {};
\node[anchor=north] at ($(l.north)+(0,-1.6mm)$) {\bfseries Law of the P\&L};
\begin{scope}[x=1mm,y=1mm,shift={($(l.center)+(-15mm,-4mm)$)}]
  \draw[grey, line width=0.4pt] (0,0) -- (30mm,0); \draw[grey, line width=0.4pt] (0,0) -- (0,11mm);
  \draw[law, line width=1.4pt] plot[smooth, domain=0.03:0.97, samples=30] ({30*\x}, {5.5+1.4*ln(\x/(1-\x))});
  \node[font=\sffamily\scriptsize, text=grey, anchor=north] at (15mm,0) {level $\tau$};
  \node[font=\sffamily\scriptsize, text=law, anchor=west] at (24mm,10mm) {$Z^\pi_\tau$};
\end{scope}
\node[anchor=south, font=\sffamily\scriptsize, text=grey] at ($(l.south)+(0,1.3mm)$) {one critic, no preference};
\node[card=es, minimum width=36mm, minimum height=22mm, above right=-10mm and 14mm of l] (e) {};
\node[anchor=north] at ($(e.north)+(0,-1.4mm)$) {\bfseries Expected shortfall};
\begin{scope}[x=1mm,y=1mm,shift={($(e.center)+(-11mm,-6mm)$)}]
  \draw[grey, line width=0.4pt] (0,0) -- (22mm,0); \draw[grey, line width=0.4pt] (0,0) -- (0,9mm);
  \draw[es, line width=1.4pt] (0,8mm) -- (5mm,8mm) -- (5mm,0) -- (22mm,0);
  \node[font=\sffamily\scriptsize, text=es, anchor=west] at (7mm,5mm) {fixed weight};
\end{scope}
\node[card=ent, minimum width=36mm, minimum height=22mm, below right=-10mm and 14mm of l] (u) {};
\node[anchor=north] at ($(u.north)+(0,-1.4mm)$) {\bfseries Entropic utility};
\begin{scope}[x=1mm,y=1mm,shift={($(u.center)+(-11mm,-6mm)$)}]
  \draw[grey, line width=0.4pt] (0,0) -- (22mm,0); \draw[grey, line width=0.4pt] (0,0) -- (0,9mm);
  \draw[ent, line width=1.4pt] plot[smooth, domain=0:22, samples=30] (\x, {8.5*exp(-\x/5)});
  \node[font=\sffamily\scriptsize, text=ent, anchor=west] at (7mm,5mm) {tilt from the law};
\end{scope}
\draw[arr] (m.east) -- (l.west);
\draw[arr] (l.east) -- (e.west);
\draw[arr] (l.east) -- (u.west);
\end{tikzpicture}}
\caption{The law of the profit and loss of a given policy is learned once, with no preference. Each preference is applied
afterwards: a spectral one as a fixed weight on the quantiles, the entropic utility as an exponential tilt
computed from the law.}
\label{fig:scheme}
\end{figure}
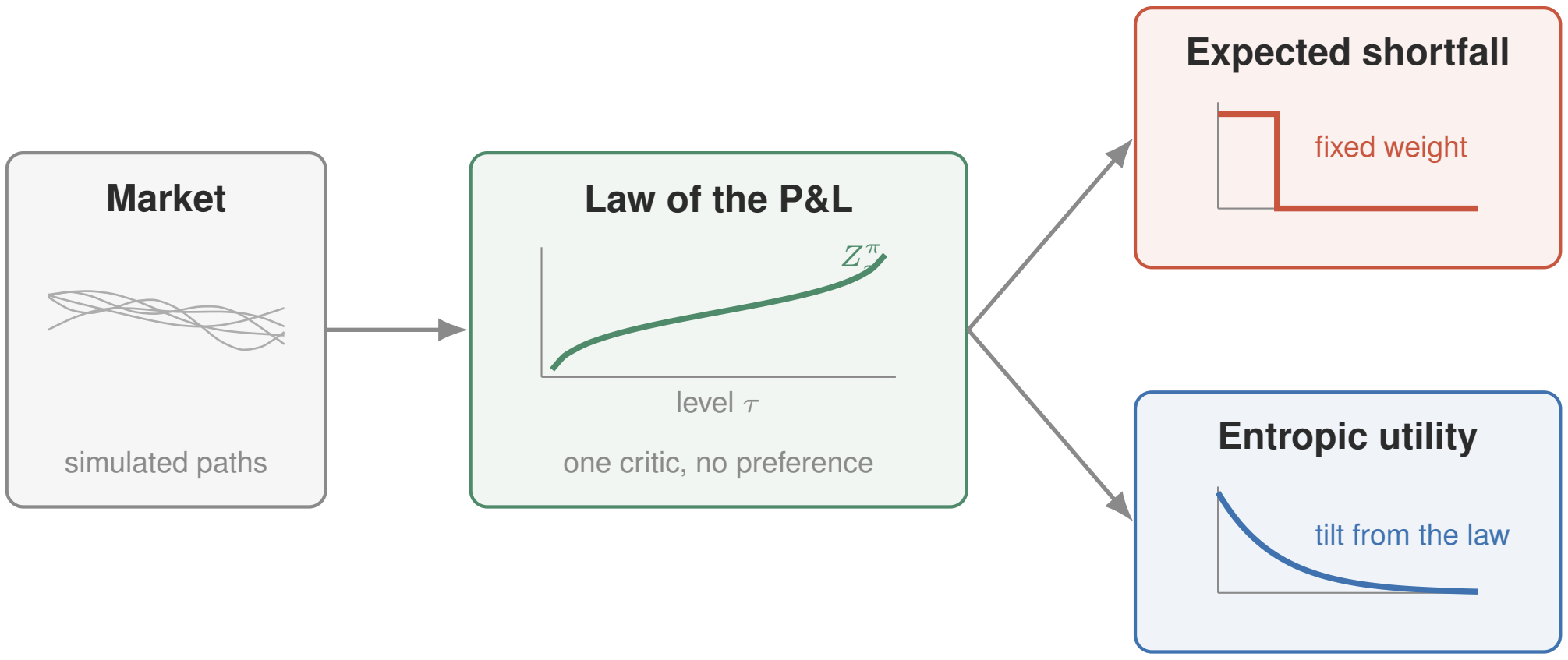

In this paper we formulate a distributional approach to deep hedging. Instead of learning a hedge, or a
value, for one preference, we learn the distribution of the profit and loss that a hedge produces, and
apply the preference afterwards. As in distributional reinforcement learning
\citep{dabney2018-quantile-regression}, the distribution of the cumulative profit and loss $Z^\pi$ is
carried by its quantile function $Z^\pi_\tau$, and a preference is a weight $w$ on that function,
\begin{equation*}
J_w(\pi) \;=\; \int_0^1 w(\tau)\, Z^\pi_\tau \, d\tau .
\end{equation*} What depends on the market and the position is then learned once, and what depends on the
preference is applied when the hedge is evaluated (Figure~\ref{fig:scheme}). This raises three questions,
which the paper answers in turn: which preferences can be applied in this way, how a hedge is optimised
once the preference is a weight on quantiles, and what the approach changes in practice.

\newpage
The paper makes three contributions.
\begin{itemize}[topsep=3pt,itemsep=3pt,parsep=0pt,leftmargin=1.4em]
\item We characterise the preferences that can be applied as a weight on the quantiles
(Section~\ref{sec:formulation} and Table~\ref{tab:preferences}). The coherent ones are exactly the
mixtures of expected shortfalls, those with a bounded weight can be read reliably from a learned
distribution, and the entropic utility, which is not of this form, is computed from the same quantiles.
Among them only the entropic utility and the mean admit a recursion on the market and the inventory, so
the problem is posed on the static objective (Section~\ref{subsec:time-consistency}).
\item We show that at the current policy each of these preferences is bounded below by an expected
utility with the same gradient (Section~\ref{sec:algorithms}). The bound turns risk-sensitive hedging
into a sequence of risk-neutral problems, and from it we derive seven algorithms: two forms of policy search,
a policy gradient, an actor--critic method, two dynamic programmes, and planning with one distributional
critic.
\item We test the algorithms against references computed by dynamic programming
(Section~\ref{sec:experiments}). The optimal hedge depends on the preference, policy search comes within
$2\%$ of the references, each preference selects its own hedge, and one critic fitted without a
preference supports planning under preferences chosen afterwards.
\end{itemize}

\begin{table}[H]
\centering
\caption{Preferences treated in the paper. Bounded weight: the score is Lipschitz in $\mathcal{W}_1$ with
constant $\lVert w\rVert_\infty$ (Proposition~\ref{prop:w1}). Cost: the factor $\mathbb{E}[\omega^2]$ of
Proposition~\ref{prop:ess}. Recursion: a Bellman equation on the market and the inventory
(Proposition~\ref{prop:time-consistency}). The power cost is infinite for $p \leq \tfrac12$.}
\label{tab:preferences}
\begin{tabular}{lllll}
\toprule
preference & weight $w(\tau)$ & bounded weight & cost $\mathbb{E}[\omega^2]$ & recursion \\
\midrule
mean & $1$ & yes & $1$ & yes \\
expected shortfall at $\alpha$ & $\alpha^{-1}\mathbf{1}\{\tau<\alpha\}$ & yes & $\alpha^{-1}$ & no \\
mixture of shortfalls & $\int_{(\tau,1]}\alpha^{-1}\mu(d\alpha)$ & if $\int\alpha^{-1}\mu(d\alpha)<\infty$ & $\iint \max(\alpha,\alpha')^{-1}\mu(d\alpha)\mu(d\alpha')$ & no \\
dual power, $q>1$ & $q(1-\tau)^{q-1}$ & yes & $q^2/(2q-1)$ & no \\
power, $\tfrac12<p<1$ & $p\tau^{p-1}$ & no & $p^2/(2p-1)$ & no \\
Wang, $a>0$ & $e^{-a\Phi^{-1}(\tau)-a^2/2}$ & no & $e^{a^2}$ & no \\
entropic, $\lambda>0$ & depends on the law & --- & depends on the law & yes \\
\bottomrule
\end{tabular}
\end{table}

The ingredients of the approach are known separately. Quantile critics and risk measures read from them
come from distributional reinforcement learning
\citep{dabney2018-quantile-regression,dabney2018-implicit-quantile-networks}, dynamic programming for
expected shortfall on an enlarged state from \citet{bauerle2011-mdp-average-value-at-risk} and
\citet{bauerle2021-spectral-risk-measures-mdp}, and policy gradients of coherent risk measures from
\citet{tamar2014-cvar-sampling,tamar2015-policy-gradient-coherent-risk}. That a preference applied date by
date does not optimise the static objective is shown by
\citet{lim2022-distributional-rl-risk-sensitive-policies} and
\citet{avilapires2025-distributional-dynamic-programming}, and the ascent from a lower bound that touches
the objective is the minorise--maximise principle \citep{hunter2004-tutorial-mm-algorithms}.
Section~\ref{sec:related-work} compares the paper with this literature. What is new is hedging with the
preference applied after training: one learned distribution serves a whole family of preferences, a
single lower bound organises the algorithms that optimise them, and the experiments show what this
changes for the hedge.

\section{Preliminaries}
\label{sec:preliminaries}

\subsection{Markov Decision Processes}
\label{subsec:mdp}

We consider the standard reinforcement learning setting in which an agent interacts with an
environment through a Markov Decision Process (MDP) \cite{sutton1998-reinforcement-learning,puterman1994-markov-decision-processes}. We
focus throughout on finite-horizon MDPs without discounting.

An undiscounted finite-horizon MDP is specified by the tuple $(\mathcal{S}, \mathcal{A},
P_{\mathcal{S}}, P_{\mathcal{R}}, g, \rho_0, T)$, where $\mathcal{S}$ denotes the state space; $\mathcal{A}$
the action space; $P_{\mathcal{S}}: \mathcal{S} \times \mathcal{A} \rightarrow
\mathscr{P}(\mathcal{S})$ the state transition kernel, mapping state-action pairs to distributions
over successor states; $P_{\mathcal{R}}: \mathcal{S} \times \mathcal{A} \rightarrow
\mathscr{P}(\mathbb{R})$ the reward kernel, mapping state-action pairs to distributions over
immediate rewards; $g: \mathcal{S} \to \mathbb{R}$ the terminal reward, received on reaching the
horizon; $\rho_0 \in \mathscr{P}(\mathcal{S})$ the initial distribution; and $T < \infty$ the
horizon itself. Here $\mathscr{P}(\cdot)$ denotes the space of
probability measures over the indicated set.

From an initial state $S_0 \sim \rho_0$, the agent at each timestep $t < T$ observes $S_t$ and selects
an action $A_t$ according to a policy $\pi: \mathcal{S} \rightarrow \mathscr{P}(\mathcal{A})$
mapping states to distributions over actions. The environment returns a reward $R_t \sim
P_{\mathcal{R}}(\cdot \mid S_t, A_t)$ and a successor state $S_{t+1} \sim P_{\mathcal{S}}(\cdot \mid
S_t, A_t)$. At $t = T$ no action is taken and the agent receives $R_T := g(S_T)$. We write $\mathbb{E}^\pi_{s,a}$ for expectation
under the law of the trajectory conditioned on $S_t = s$ and $A_t = a$, $\mathbb{E}^\pi_{s}$ for
the same conditioned on $S_t = s$, and $\mathbb{E}^\pi$ for the unconditional expectation, which
integrates over $\rho_0$ as well. Where no policy is involved the superscript is dropped, so that
$\mathbb{E}_{s,a}$ denotes expectation over $R_t$ and $S_{t+1}$ under the two kernels at $(s,a)$.

The return from timestep $t$ is the total reward collected thereafter,
$$Z^\pi_{t:T} := \sum_{t'=t}^{T-1} R_{t'} \;+\; R_T,$$
and the standard objective is to maximise its expectation,
$$J(\pi) := \mathbb{E}^\pi\big[Z^\pi_{0:T}\big].$$

Value functions summarise the return by its mean, and under a finite horizon they depend on the
timestep as well as on the state:
$$Q^\pi_t(s, a) := \mathbb{E}^\pi_{s,a}\big[Z^\pi_{t:T}\big], \quad t = 0, \dots, T-1,
\qquad
V^\pi_t(s) := \mathbb{E}^\pi_{s}\big[Z^\pi_{t:T}\big], \quad t = 0, \dots, T,$$
where $Q^\pi_t$ evaluates an action taken at $s$ with $\pi$ followed thereafter and $V^\pi_t$
evaluates $s$ under $\pi$ \cite{watkins1992-q-learning,bertsekas2022-abstract-dynamic-programming}. Additivity of the return and the tower
property give the Bellman equations
$$Q^\pi_t(s,a) = \mathbb{E}^\pi_{s,a}\big[R_t + V^\pi_{t+1}(S_{t+1})\big], \qquad
V^\pi_t(s) = \mathbb{E}_{a \sim \pi(\cdot \mid s)}\big[Q^\pi_t(s,a)\big].$$
A policy $\pi^*$ is optimal if $V^{\pi^*}_t(s) \geq V^\pi_t(s)$ for every $t$, every $s \in
\mathcal{S}$ and every $\pi$; writing $V^*$ and $Q^*$ for its value functions, these satisfy the
Bellman optimality equations
$$Q^*_t(s,a) = \mathbb{E}_{s,a}\big[R_t + V^*_{t+1}(S_{t+1})\big], \qquad
V^*_t(s) = \max_{a \in \mathcal{A}} Q^*_t(s,a),$$
in which the averaging over the policy's own action is replaced by a maximisation over actions.
Both recursions are initialised at the horizon, where $V^\pi_T \equiv V^*_T \equiv g$.

We work throughout with deep reinforcement learning, in which the policy and any value function
are represented by neural networks and trained by stochastic gradient descent
\cite{mnih2015-human-level-control,silver2016-mastering-go}. Two families of algorithm are relevant
here. Policy search methods restrict attention to a parametrised family of policies $\pi_\theta$
and optimise $J(\pi_\theta)$ over $\theta$ directly, either along an estimate of $\nabla_\theta J$
obtained from sampled trajectories or by derivative-free search, without maintaining a value
function at all. Actor-critic methods \cite{konda1999-actor-critic,mnih2016-asynchronous-methods}
maintain both: a critic $Q_{\phi,t} \approx Q^{\pi_\theta}_t$ is fitted to the return of the
current policy, and the actor $\pi_\theta$ is then updated using the critic, for a deterministic
policy along $\nabla_\theta J \approx \mathbb{E}^{\pi_\theta}\big[\sum_{t<T} \nabla_\theta
Q_{\phi,t}(S_t, \pi_\theta(S_t))\big]$ \citep{silver2014deterministic-policy-gradient,lillicrap2019-continuous-control}, the critic
being held fixed so that $\theta$ is differentiated only through the action.

The framework above evaluates a policy through the mean of its return. Distributional
reinforcement learning \citep{bellemare-2017-c51-distributional-rl,rowland2018-analysis-categorical-distributional-rl,%
dabney2018-quantile-regression,dabney2018-implicit-quantile-networks,bellemare2023-distributional-rl}
learns the distribution of the return directly. Write $Z^\pi_{t:T}(s,a)$ and $Z^\pi_{t:T}(s)$ for the
return conditioned on $(S_t, A_t) = (s,a)$ and on $S_t = s$ respectively, so that $Q^\pi_t$ and
$V^\pi_t$ are their means. The object estimated is the map $\mathcal{S} \times \mathcal{A} \to
\mathscr{P}(\mathbb{R})$ sending $(s,a)$ to the law of $Z^\pi_{t:T}(s,a)$, in place of the value
function $Q^\pi_t: \mathcal{S} \times \mathcal{A} \to \mathbb{R}$. It satisfies the distributional
Bellman equation
\begin{equation}
\label{eq:dist-bellman}
Z^\pi_{t:T}(s,a) \;\overset{d}{=}\; R_t + Z^\pi_{t+1:T}(S_{t+1}), \qquad
S_{t+1} \sim P_{\mathcal{S}}(\cdot \mid s,a), \qquad Z^\pi_{T:T}(s) = g(s),
\end{equation}
an identity between laws rather than between random variables. Functionals of the return other
than its expectation then become accessible.

The following subsection formalises the hedging problem within this MDP structure, specifying the
state representation, action space and reward function appropriate for derivatives trading.

\subsection{Trading Environments as Markov Decision Processes}
\label{subsec:trading-mdp}

We formulate the derivatives hedging problem as a finite-horizon MDP following
\citet{buehler2019-deep-hedging}. An agent holds a portfolio of contingent claims and trades $d_h$
liquid hedging instruments over $T$ dates in order to control the risk of that portfolio. The
formulation captures market observables, portfolio dynamics, transaction costs and contract
maturities without committing to any particular model of price dynamics.

The state space factorises into a component the agent does not control and a component recording
the consequences of its own decisions, $\mathcal{S} = \mathcal{X} \times \mathcal{Y}$ with
$\mathcal{Y} = \mathbb{R}^{d_h} \times \mathbb{R}$, and we write
\[
S_t \;=\; (X_t,\, Y_t), \qquad Y_t \;=\; (\Delta_t,\, W_t).
\]
The \emph{market state} $X_t$ collects the observable market variables, namely the prices of the
hedging instruments and their implied volatilities; whatever else the price model requires in order
to be Markov, which is nothing beyond the spot under geometric Brownian motion and the
instantaneous variance under Heston; and the contingent-claim book, namely the liability being
hedged together with any path-dependent information its payoff requires, such as a barrier-breach
indicator. The \emph{agent state} $Y_t$ collects the inventory $\Delta_t$ held in the $d_h$
instruments and the profit and loss $W_t$ accumulated so far, with $\Delta_0$ the initial inventory
and $W_0 = 0$.

At $t < T$ the agent selects a trade $A_t \in \mathcal{A}_t(S_t) \subseteq \mathbb{R}^{d_h}$, the
number of units bought in each instrument, negative entries denoting sales, where the admissible
set may encode state-dependent constraints such as position limits or liquidity bounds. The
dynamics have three components.

\begin{itemize}[topsep=3pt,itemsep=3pt,parsep=0pt,leftmargin=2.9em]
\item[\textnormal{(E1)}] \textbf{No market impact.} The market state evolves on its own,
$X_{t+1} \sim P^{\mathrm{mkt}}_t(\cdot \mid X_t)$, a kernel on $\mathcal{X}$ that takes no action
argument: the agent is a price taker.

\item[\textnormal{(E2)}] \textbf{Rewards are cash flows.} Let $H_t(X_t) \in \mathbb{R}^{d_h}$ be
the mid-prices of the hedging instruments, $D_t(X_t)$ the cash the book generates at date $t$,
comprising claim payoffs together with any coupons or dividends on the hedge, and
$c_t(X_t, A_t) \geq 0$, with $c_t(X_t, \mathbf{0}) = 0$, a friction covering bid--ask spreads and
fees. Trading $A_t$ costs $A_t^\top H_t(X_t)$ at mid-prices plus that friction, so the reward is
the net cash received,
\[
R_t \;=\; D_t(X_t) - A_t^\top H_t(X_t) - c_t(X_t, A_t), \qquad t < T,
\]
and at the horizon the book settles and the inventory is liquidated,
\[
R_T \;=\; g(S_T) \;=\; D_T(X_T) + \Delta_T^\top H_T(X_T) - c_T(X_T, -\Delta_T).
\]
The first is a deterministic function of the state and the action, so $P_{\mathcal{R}}$ places
all of its mass at a point; the second is the terminal reward $g$ of Section~\ref{subsec:mdp}.

\item[\textnormal{(E3)}] \textbf{The agent state accumulates.} The inventory and the profit and
loss are the running sums of the trades and of the rewards,
\begin{equation}
\label{eq:accumulator}
Y_{t+1} \;=\; Y_t + (A_t,\, R_t),
\qquad\text{that is}\qquad
\Delta_t = \Delta_0 + \sum_{k<t} A_k, \quad W_t = \sum_{k<t} R_k .
\end{equation}
\end{itemize}

The terminal reward adopts the \emph{liquidation} convention, charging the friction of closing the
book; marking the inventory to market at $H_T$ without that charge defines a different problem, and
values obtained under the two conventions are not comparable.

Two properties of this environment are used in what follows. The first is that the profit and loss
is determined by the terminal state: summing \eqref{eq:accumulator} gives
\begin{equation}
\label{eq:pnl-is-W}
Z^\pi \;=\; Z^\pi_{0:T} \;=\; W_T + R_T,
\end{equation}
in which $W_T$ is a coordinate of $S_T$ and $R_T = g(S_T)$ a function of it, so any objective
taken over the profit and loss is a functional of the terminal state.

The second is that nothing is lost by restricting to policies. A hedging strategy is classically an
adapted process, a function of the entire market history, whereas a policy is a function of $S_t$
alone; under (E1) the entropic problem and every finite shortfall mixture admit optimal policies of $S_t$
whenever the one-step maxima are attained (Propositions~\ref{prop:es-dp} and~\ref{prop:entropic-dp}),
and a general spectrum does as well \citep{bauerle2021-spectral-risk-measures-mdp}. The inventory alone would not always
suffice, since two paths reaching the same $\Delta_t$ at the same $X_t$ may have paid different
amounts to do so: $\Delta_t$ records the sum of the trades and $W_t$ the prices at which they were
made. Which objectives need $W_t$ is the question of Section~\ref{subsec:time-consistency}: the
entropic recursion runs on $(X_t, \Delta_t)$, called below the \emph{market-and-inventory state},
and a static spectral objective needs $S_t$ itself, called the \emph{enlarged state}.

This environment is the same for a risk-neutral hedger as for a risk-averse one. What separates
them is the objective taken over the profit and loss, which the following subsection states in the
form the deep hedging literature uses.

\subsection{Deep Hedging}
\label{subsec:deep-hedging}

The Deep Hedging framework, introduced by \citet{buehler2019-deep-hedging}, reformulates
derivatives hedging as the direct optimisation of a risk-adjusted performance measure. Deep Hedging
learns a policy in the environment of Section~\ref{subsec:trading-mdp}, optimising for an
institution's risk preferences while explicitly accounting for transaction costs, discrete trading
dates and market incompleteness.

The key innovation is the use of a utility function to encode risk preferences. Let $Z^\pi$ denote
the cumulative profit and loss obtained by following policy $\pi$ over the hedging horizon, given
by \eqref{eq:pnl-is-W}. A utility function $\mathcal{U}: \mathscr{P}(\mathbb{R}) \rightarrow
\mathbb{R}$ maps the distribution of $Z^\pi$ to a scalar representing the desirability of that
distribution under a specific risk attitude; being a function of the law alone, it is written
$\mathcal{U}(X)$ for a random variable $X$ without ambiguity.
\begin{problem}[Deep Hedging]
\label{prob:dh}
Let $(\mathcal{S}, \mathcal{A}, P_{\mathcal{S}}, P_{\mathcal{R}}, g, \rho_0, T)$ denote the hedging MDP of
Section~\ref{subsec:trading-mdp}, and let $\mathcal{U}: \mathscr{P}(\mathbb{R}) \rightarrow
\mathbb{R}$ be a risk-adjusted utility function representing institutional preferences. The Deep
Hedging problem seeks an optimal policy $\pi^*: \mathcal{S} \rightarrow \mathscr{P}(\mathcal{A})$
maximising the utility of the cumulative profit and loss:
\[
\pi^* \in \argmax_{\pi} J_{\mathcal{U}}(\pi), \qquad \text{where} \qquad J_{\mathcal{U}}(\pi) :=
\mathcal{U}(Z^\pi).
\]
It covers the risk-neutral objective, $\mathcal{U}(X) = \mathbb{E}[X]$, and risk-averse
objectives over the tail and the higher moments of the profit and loss.
\end{problem}

To ensure economically coherent strategies, utility functions are typically required to satisfy
structural properties \citep{artzner1999-coherent-measures-of-risk}. Monotonicity requires $\mathcal{U}(X) \geq
\mathcal{U}(Y)$ when $X \geq Y$ almost surely. Concavity encodes risk aversion through
$\mathcal{U}(\kappa X + (1-\kappa)Y) \geq \kappa \mathcal{U}(X) + (1-\kappa)\mathcal{U}(Y)$ for
$\kappa \in [0,1]$. Cash invariance stipulates $\mathcal{U}(X + c) = \mathcal{U}(X) + c$ for
constants $c \in \mathbb{R}$, so that the score is quoted in the units of the profit and loss and
the premium received for writing the book shifts it by a constant. Normalisation sets
$\mathcal{U}(0) = 0$.

Several utility functions have been employed in the Deep Hedging literature:

\begin{itemize}[topsep=3pt,itemsep=3pt,parsep=0pt,leftmargin=1.4em]
\item \textbf{Expectation:} $\mathcal{U}(X) = \mathbb{E}[X]$, risk-neutral optimisation, maximising
average profit without regard for variability or tail risk.

\item \textbf{Expected Shortfall:} for a tail fraction $\alpha \in (0,1]$, the average of the
worst $\alpha$ fraction of outcomes, widely used in regulation. It is not the expectation of a fixed function of $X$, such a functional being affine in
the law of $X$ where expected shortfall is concave, but it is one optimised over a single scalar,
\begin{equation}
\label{eq:es}
\mathcal{U}_\alpha(X) \;=\; \sup_{b \in \mathbb{R}} \Big\{ b - \frac{1}{\alpha}
\mathbb{E}\big[(b - X)^+\big] \Big\},
\end{equation}
attained at any $\alpha$-quantile of $X$, that is any $b^\star$ with $\mathbb{P}[X < b^\star] \leq
\alpha \leq \mathbb{P}[X \leq b^\star]$
\citep{rockafellar2000-optimization-of-cvar,rockafellar2002-cvar-general-loss-distributions}. It is
coherent \citep{artzner1999-coherent-measures-of-risk,acerbi2002-coherence-expected-shortfall}, hence
positively homogeneous. Average value at risk and conditional value at
risk name the same functional; the tail conditional expectation $\mathbb{E}[X \mid X \leq
b^\star]$ does not, once $\mathbb{P}[X \leq b^\star] > \alpha$, as for a discretely rebalanced
hedge \citep{acerbi2002-coherence-expected-shortfall}.

\item \textbf{Entropic Utility:} the certainty equivalent of the exponential utility
$u_\lambda(x) = -e^{-\lambda x}$, that is the deterministic amount an agent with that utility
would accept in place of $X$, where $\lambda > 0$ is the risk aversion:
\begin{equation}
\label{eq:entropic}
\mathcal{U}_\lambda(X) = -\frac{1}{\lambda} \log \mathbb{E}\big[\exp(-\lambda X)\big].
\end{equation}
When $X$ has finite exponential moments, $\mathcal{U}_\lambda(X) = \mathbb{E}[X] -
\tfrac{\lambda}{2}\operatorname{Var}(X) + o(\lambda)$ as $\lambda \downarrow 0$, which is the
mean--variance reading of the parameter. It approaches $\mathbb{E}[X]$ as $\lambda \to 0$ and $\operatorname*{ess\,inf} X$ as $\lambda \to
\infty$ \citep{kupper2009-law-invariant-time-consistent}, so a single parameter spans risk
neutrality to the worst case. Unlike expected shortfall it is not positively homogeneous, since
doubling a position does not double its entropic value, so it is a concave monetary utility rather than a
coherent one, a convex risk measure in the sign convention of
\citet{follmer2010-convex-coherent-risk-measures}. It
is, however, time consistent, and therefore decomposes across dates
\citep{kupper2009-law-invariant-time-consistent}, which is what allows the Bellman treatment
below.
\end{itemize}

Algorithms for Problem~\ref{prob:dh} fall into two groups. Policy search represents the policy as
a neural network $\pi_\theta$ and ascends $J_{\mathcal{U}}(\pi_\theta)$ directly on simulated
market paths, the gradient obtained by backpropagation through the rollout; this is the original
formulation \citep{buehler2019-deep-hedging}, and it maintains no value function, so the learning
signal reaches every trade only through the terminal profit and loss, and it is the route most of
the literature follows \citep{carbonneau2020-deep-hedging-long-term-derivatives,imaki2021-no-transaction-band-network}.
Actor-critic formulations add one, deriving a risk-averse Bellman representation for utilities that
decompose across dates and fitting a critic against it, as in \citet{murray2022-deep-hedging-continuous-rl}
for the entropic utility and \citet{buehler2022-deep-bellman-hedging} for monetary utilities generally;
critics also track the moments of the hedging cost \citep{cao2021-deep-hedging-reinforcement-learning},
its quantiles \citep{cao2023-gamma-vega-hedging} or a recursive expectile
\citep{marzban2023-deep-rl-expectile-risk}. In the
first group the objective is defined relative to a fixed $\mathcal{U}$; in the second the critic
is as well. The following section learns the law of the profit and
loss first and applies the preference afterwards.

\section{Distributional Deep Hedging}
\label{sec:formulation}

\subsection{Problem Formulation}
\label{subsec:ddh}

The formulation of Section~\ref{subsec:deep-hedging} specifies a utility before training and
learns a value function defined relative to it. A single position, however, is ordinarily
evaluated under several risk preferences at once: regulatory capital is reported as expected
shortfall at prescribed levels, internal limits combine such levels in fixed proportions, and a
partially hedged position must be valued at intermediate dates. Each such evaluation requires a
separately trained object whenever the preference is embedded in the quantity being learned.

The key change is to take the distribution of the profit and loss as the primitive object and to
recover the preference as a separate choice made afterwards. Let $Z^\pi$ denote the cumulative
profit and loss of a policy $\pi$, given by \eqref{eq:pnl-is-W}, and $F_{Z^\pi}$ its distribution
function. The object learned is the law of $Z^\pi$, carried by its quantile function
\[
Z^\pi_\tau := F^{-1}_{Z^\pi}(\tau) = \inf\big\{x \in \mathbb{R} : F_{Z^\pi}(x) \geq \tau\big\},
\qquad \tau \in (0,1),
\]
the generalised inverse being used so that the definition survives the atoms a discretely
rebalanced position produces. The representation is complete: if $\tau \sim U(0,1)$ then
$Z^\pi_\tau \overset{d}{=} Z^\pi$. A preference is then a rule for scoring that function, and we
take the scoring to be linear in it,
\begin{equation}
\label{eq:interface}
J_w(\pi) \;:=\; \int_0^1 w(\tau)\, Z^\pi_\tau \, d\tau,
\end{equation}
where the weight $w: (0,1) \to [0,\infty)$, with $\int_0^1 w = 1$, carries the preference and nothing else;
for a random variable $X$ with finite first moment, $J_w(X) := \int_0^1 w\,F^{-1}_X$, so $J_w(\pi) = J_w(Z^\pi)$. Under the
conditions of Section~\ref{subsec:admissible} the functionals of this form are the \emph{spectral}
risk measures and $w$ is the \emph{risk spectrum}
\citep{acerbi2002-spectral-measures-of-risk}. A policy is thus scored by a weighted average of its
possible outcomes, $w$ determining how much each one counts. The quantile function
depends on the policy and the environment alone, the weight on the preference alone, and the two
meet only in the integral. A critic fitted to $Z^\pi_\tau$ therefore carries no risk parameter, and
one such critic scores every preference in the family, at every state, without retraining.

\begin{problem}[Distributional Deep Hedging]
\label{prob:ddh}
Let $(\mathcal{S}, \mathcal{A}, P_{\mathcal{S}}, P_{\mathcal{R}}, g, \rho_0, T)$ denote the hedging
MDP of Section~\ref{subsec:trading-mdp}, and let $w$ be a risk spectrum. The Distributional Deep
Hedging problem seeks a policy $\pi^* \in \argmax_{\pi} J_w(\pi)$, maximising the weighted
quantile average \eqref{eq:interface} of the cumulative profit and loss.
\end{problem}

The preference is often carried as a \emph{distortion} rather than as a weight. The distortion is
the antiderivative of the weight,
\begin{equation}
\label{eq:distortion}
\beta(\tau) \;:=\; \int_0^\tau w(u)\,du,
\qquad\text{equivalently}\qquad
w \;=\; \beta' ,
\end{equation}
so $\beta$ is non-decreasing with $\beta(0)=0$ and $\beta(1)=1$, and is concave exactly when $w$ is
non-increasing. Then $J_w(\pi) = \int_0^1 Z^\pi_\tau\,d\beta(\tau)$, and changing variables
through the generalised inverse removes the weight from the integrand,
\begin{equation}
\label{eq:distorted-sampling}
J_w(\pi) \;=\; \mathbb{E}_{u \sim U(0,1)}\big[\,Z^\pi_{\beta^{-1}(u)}\,\big],
\end{equation}
which is the form used in practice: levels are drawn uniformly and distorted through the generalised inverse $\beta^{-1}$,
so the critic is evaluated only where $w$ places weight. Several distortions are in common use:

\begin{itemize}[topsep=3pt,itemsep=3pt,parsep=0pt,leftmargin=1.4em]
\item \textbf{Identity}, $\beta(\tau) = \tau$: uniform weight on every outcome, recovering the
risk-neutral objective.

\item \textbf{Expected shortfall} at tail fraction $\alpha \in (0,1]$: weight flat on the worst
$\alpha$ fraction and zero elsewhere, so that $\beta^{-1}(u) = \alpha u$ and every drawn level
lands inside the tail:
$$\beta_\alpha(\tau) = \min(\tau/\alpha,\,1) .$$

\item \textbf{Dual power}, $q > 1$: a smooth preference whose weight is bounded, equal to $q$ at
the worst outcome and vanishing at the best:
$$\beta(\tau) = 1-(1-\tau)^{q} .$$

\item \textbf{Power}, $\beta(\tau) = \tau^{p}$ with $0 < p < 1$: smooth in the same way, but with
weight \emph{unbounded} at the worst outcome.

\item \textbf{Wang transform}, $a > 0$: introduced in actuarial science
\citep{wang2000-class-of-distortion-operators}, it reweights the whole distribution rather than
only the tail, and is also unbounded at the worst outcome:
$$\beta_a(\tau) = \Phi\big(\Phi^{-1}(\tau)+a\big),$$
where $\Phi$ is the standard normal distribution function.
\end{itemize}

The last two place unbounded weight near $\tau = 0$, so $\beta'$ is unbounded at the origin;
both are in common use, and Section~\ref{subsec:estimability} examines the consequence.
\subsection{Coherent Preferences and Expected Shortfall}
\label{subsec:admissible}\label{subsec:es}

Problem~\ref{prob:ddh} is posed for an arbitrary weight. This subsection states which weights a risk measure is required to have, and shows that they are exactly the mixtures of expected shortfalls.

A functional $\rho$ is \emph{coherent} \citep{artzner1999-coherent-measures-of-risk,follmer2010-convex-coherent-risk-measures} when it has
four properties, which in the convention used here, where larger is better, are:
\emph{monotonicity}, $\rho(X) \geq \rho(Y)$ whenever $X \geq Y$ almost surely; \emph{cash
invariance}, $\rho(X+c) = \rho(X)+c$, so the score is quoted in the units of the outcome;
\emph{positive homogeneity}, $\rho(cX) = c\,\rho(X)$ for $c>0$; and \emph{superadditivity},
$\rho(X+Y) \geq \rho(X)+\rho(Y)$, so that combining two positions never scores worse than holding
them apart.

For $J_w$ each of these is a condition on the weight. Positive homogeneity is automatic, since
$J_w$ is linear in the quantile function and scaling an outcome scales every quantile. The other
three hold exactly when
\begin{equation}
\label{eq:coherence-conditions}
w \;\geq\; 0, \qquad \int_0^1 w \;=\; 1, \qquad w \ \text{non-increasing},
\end{equation}
giving monotonicity, cash invariance and superadditivity respectively, up to null sets
\citep{acerbi2002-spectral-measures-of-risk}. The third expresses risk
aversion: a weight decreasing in $\tau$ charges bad outcomes more heavily than good ones.

\begin{proposition}[Order properties]
\label{prop:order}
Let $w \geq 0$ with $\int_0^1 w = 1$. Then $J_w$ is monotone with respect to first-order
stochastic dominance, and additive on comonotone pairs: if $X$ and $Y$ are non-decreasing
functions of a common random variable, $J_w(X+Y) = J_w(X) + J_w(Y)$. If $w$ is also
non-increasing, $J_w$ is monotone with respect to the concave order where it is finite, so a
mean-preserving spread cannot raise the score.
\end{proposition}

\begin{proof}
Dominance is $F^{-1}_X \geq F^{-1}_Y$ pointwise, which $w \geq 0$ preserves in
\eqref{eq:interface}; comonotonicity gives $F^{-1}_{X+Y} = F^{-1}_X + F^{-1}_Y$, and
\eqref{eq:interface} is linear in the quantile function. For the concave order put
$\Psi_X(t) = \int_0^t F^{-1}_X$, so a mean-preserving spread means $\Psi_Y \leq \Psi_X$ with
equality at $t = 1$. Integrating by parts, the boundary terms vanish, since $\Psi_X - \Psi_Y$ is
zero at both ends, leaving, for bounded $w$,
$J_w(X) - J_w(Y) = -\int_0^1 (\Psi_X - \Psi_Y)\,dw \geq 0$, since $dw \leq 0$; for unbounded $w$
apply this to $\min(w, n)$ and let $n \to \infty$ by monotone convergence.
\end{proof}

Comonotone additivity is the case of equality in superadditivity: two books moving together offer
no diversification to credit. It also explains why \eqref{eq:interface} is not a modelling choice.
\citet{kusuoka2001-law-invariant-coherent-risk-measures} shows that on an atomless space every
law-invariant coherent functional with the Fatou property is an infimum of mixtures of expected shortfalls, and that
comonotone additivity collapses that infimum to a single mixture, namely an integral against a
concave distortion, which is \eqref{eq:interface}. The linear form is therefore forced once comonotone additivity is required in addition to law invariance and coherence.

The theorem fixes the form of the objective but describes the weights
that result only implicitly. The following representation describes them directly, and identifies
expected shortfall $\mathcal{U}_\alpha$ of \eqref{eq:es}, the average of the worst $\alpha$ fraction of outcomes,
as the generator of the admissible family.

\begin{proposition}[Mixture representation; \citealp{kusuoka2001-law-invariant-coherent-risk-measures,acerbi2002-spectral-measures-of-risk}]
\label{prop:mixture}
A function $w:(0,1) \to [0,\infty)$ is non-increasing with $\int_0^1 w = 1$ if and only if there
is a probability measure $\mu$ on $(0,1]$ with
$w(\tau) = \int_{(\tau,1]} \alpha^{-1}\mu(d\alpha)$ for almost every $\tau$. In that case, for $X$ with finite first moment,
\[
J_w(X) \;=\; \int_{(0,1]} \mathcal{U}_\alpha(X)\, \mu(d\alpha),
\qquad
\lVert w \rVert_\infty \;=\; \int_{(0,1]} \alpha^{-1}\, \mu(d\alpha) .
\]
\end{proposition}

\begin{proof}
Given such a $\mu$, Tonelli gives $\int_0^1 w = \int_{(0,1]}\alpha^{-1}\!\int_0^1
\mathbf{1}\{\tau<\alpha\}\,d\tau\,\mu(d\alpha) = 1$, and $w$ is non-increasing by inspection.
Conversely take the right-continuous version of $w$ and put $\mu(d\alpha) = \alpha\,d(-w)(\alpha)$ on $(0,1)$ with an atom $w(1^-)$ at $1$; integrating by parts, since $\alpha w(\alpha) \leq \beta(\alpha) \to 0$, its mass is $\int_0^1 w = 1$. Fubini gives
$\int \mathcal{U}_\alpha(X)\mu(d\alpha) = \int_0^1 F^{-1}_X(\tau) w(\tau)\,d\tau$, and the norm is
$w(0^+)$.
\end{proof}

Every admissible objective is therefore an average of expected shortfalls, and every such average
is admissible; $\mathcal{U}_\alpha$ is the case $\mu = \delta_\alpha$ and the mean is
$\mu = \delta_1$. Choosing a coherent preference is choosing how to distribute attention across
depths of tail, and $\mu$ is that choice. It is also the form in which such a choice is usually
stated: a capital charge at a prescribed level names a single $\alpha$, and an internal limit
combining levels names the mass it puts on each.

A preference stated at finitely many levels is a measure with finitely many atoms. For
$\mu = \sum_{k=1}^K c_k\,\delta_{\alpha_k}$ with $\alpha_1 < \dots < \alpha_K$, $c_k > 0$ and
$\sum_k c_k = 1$, the weight is the step function $w(\tau) = \sum_{k:\,\alpha_k > \tau} c_k/\alpha_k$,
constant between consecutive levels and zero above $\alpha_K$; the objective is
$\sum_k c_k\,\mathcal{U}_{\alpha_k}$ and $\lVert w \rVert_\infty = \sum_k c_k/\alpha_k$.

\subsection{Relation to the Entropic Utility}
\label{subsec:entropic}

The entropic utility \eqref{eq:entropic}, on which deep hedging is customarily trained, is not a
member of the family. Every $J_w$ is positively homogeneous, $J_w(cX) = c\,J_w(X)$ for $c > 0$,
being linear in the quantile function, whereas $\mathcal{U}_\lambda$ is not; no choice of $w$
reproduces it.

The following identity measures the difference. Scaling inside the entropic utility can be traded
against the risk aversion,
\begin{equation}
\label{eq:entropic-scaling}
\mathcal{U}_\lambda(cX) \;=\; c\,\mathcal{U}_{c\lambda}(X), \qquad c > 0,
\end{equation}
since both sides equal $-\lambda^{-1}\log\mathbb{E}[e^{-\lambda c X}]$. Doubling a position
therefore doubles the score only at a doubled risk aversion, whereas positive homogeneity would double it at the same one. A spectral
objective carries no parameter able to absorb the scaling, so the two families agree only in the
limit $\lambda \downarrow 0$.

The separation of Section~\ref{subsec:ddh} nonetheless holds for the gradient, which a
risk-agnostic critic supplies exactly.

\begin{proposition}[Entropic gradient from a risk-agnostic critic]
\label{prop:entropic}
Fix a state and an action $a$, write $Z_\tau(a)$ for the quantile function of the resulting
return, and suppose $\int_0^1 e^{-\lambda Z_\tau(a)}\,d\tau$ is finite and positive with
$\lambda\lVert\nabla_a Z_\tau\rVert e^{-\lambda Z_\tau}$ dominated near $a$. Then
\[
\mathcal{U}_\lambda \;=\; -\frac{1}{\lambda}\log \int_0^1 e^{-\lambda Z_\tau(a)}\,d\tau,
\qquad
\nabla_a\,\mathcal{U}_\lambda \;=\; \int_0^1 w^\star(\tau;a)\,\nabla_a Z_\tau(a)\,d\tau,
\]
where the weight is the normalised exponential
\[
w^\star(\tau;a) \;=\; \frac{e^{-\lambda Z_\tau(a)}}{\int_0^1 e^{-\lambda Z_{\tau'}(a)}\,d\tau'} .
\]
\end{proposition}

\begin{proof}
Write $\Lambda(a) = \int_0^1 e^{-\lambda Z_\tau(a)}\,d\tau$. Since $\tau \sim U(0,1)$ reproduces
the law, $\mathbb{E}[e^{-\lambda Z}] = \Lambda(a)$, and applying $-\lambda^{-1}\log$ gives the
value. Domination is the hypothesis of Leibniz's rule, so differentiating under the integral sign
gives
\[
\nabla_a\Lambda(a) \;=\; -\lambda\int_0^1 e^{-\lambda Z_\tau(a)}\,\nabla_a Z_\tau(a)\,d\tau,
\]
and $\nabla_a\mathcal{U}_\lambda = -\lambda^{-1}\nabla_a\Lambda/\Lambda$, which on dividing by
$\Lambda$ is the stated integral.
\end{proof}

The entropic \emph{gradient} is thus a weighted average of the critic's own action-gradients, the
weights being a softmax of its outputs multiplied by $-\lambda$, and the density of an
exponential tilt of the return distribution. Only the origin of the weight differs: a spectral
preference fixes the attention each \emph{rank} receives before anything is learned, while the entropic
utility computes $w^\star$ from the learned law, fixing the attention paid to each \emph{value}.

The \emph{value}, by contrast, is not of the form \eqref{eq:interface}. By the Gibbs variational
principle \citep{dembo2010-large-deviations} the two differ by
\begin{equation}
\label{eq:kl-gap}
\mathcal{U}_\lambda \;-\; \int_0^1 w^\star(\tau)\,Z_\tau\,d\tau
\;=\; \tfrac{1}{\lambda}\,D_{\mathrm{KL}}\big(w^\star \,\Vert\, 1\big) \;\geq\; 0,
\end{equation}
which vanishes only as $\lambda \downarrow 0$ or when $Z$ is constant. When $Z$ has finite exponential
moments the gap is $\tfrac{\lambda}{2}\operatorname{Var}(Z) + O(\lambda^2)$ as $\lambda \downarrow 0$. One risk-agnostic critic therefore serves both
families, but only for a spectral preference is the objective a weighted quantile average; the entropic
value is reported through the logarithm of Proposition~\ref{prop:entropic}.

The prevalence of the entropic utility in this literature follows from a classification result.
On a standard filtered probability space, \citet[Theorem~1.10]{kupper2009-law-invariant-time-consistent}
show that the dynamic risk measures, which are cash invariant by definition, that are law
invariant, relevant and time consistent (Section~\ref{subsec:time-consistency}) are exactly the
entropic family \eqref{eq:entropic}, with $\lambda \in (-\infty, \infty]$, read as the mean at $\lambda = 0$ and the essential infimum at $\lambda = \infty$.

The entropic utility is finite only for laws with exponential moments, and the two families part there
too. Under the price models of Section~\ref{sec:experiments}, a short call hedged with fewer than one share
over an interval of positive probability has $\mathbb{E}[e^{-\lambda Z}] = \infty$ and entropic utility
$-\infty$, whereas a spectral objective with bounded weight needs only a finite first moment. As in deep
hedging \citep{buehler2019-deep-hedging}, the problems are therefore posed on finite probability spaces:
the simulated paths in training and evaluation, and the quadrature lattice in the programmes. On these every
profit and loss is bounded and every proposition of the paper applies.

The two families differ again in the algorithms they admit. The entropic
utility is time consistent \citep{kupper2009-law-invariant-time-consistent,%
cheridito2009-time-inconsistency-of-var,ruszczynski2010-risk-averse-dynamic-programming}, and no spectral objective other than the mean is time consistent on every filtration; Section~\ref{subsec:time-consistency} states what each requires of an algorithm.

\subsection{Time Consistency}
\label{subsec:time-consistency}

A hedge is revised at every rebalancing date, and a procedure that improves it one date at a time
presumes that what the preference favours from a later date it also favours from inception. Write
$\rho_t(X)$ for a preference applied to the law of $X$ conditional on $\mathcal{F}_t$, the
information at date $t$. The family $(\rho_t)$ is \emph{time consistent} when
$\rho_{t+1}(X) \geq \rho_{t+1}(Y)$ almost surely implies $\rho_t(X) \geq \rho_t(Y)$; for
cash-invariant, monotone families this is the recursion $\rho_t(X) = \rho_t(\rho_{t+1}(X))$
\citep{cheridito2009-time-inconsistency-of-var,ruszczynski2010-risk-averse-dynamic-programming}, which
is what a Bellman equation on the market-and-inventory state requires. The proposition below is stated on
$L^\infty$, the example after it on square-integrable increments.

\begin{proposition}[Time consistency in the two families]
\label{prop:time-consistency}
The entropic family is time consistent,
$\mathcal{U}_\lambda(X \mid \mathcal{F}_t) = \mathcal{U}_\lambda\big(\mathcal{U}_\lambda(X \mid
\mathcal{F}_{t+1}) \,\big|\, \mathcal{F}_t\big)$. A coherent spectral preference applied
conditionally is time consistent on every filtration only if it is the mean, $w \equiv 1$.
\end{proposition}
\begin{proof}
For the first statement, $\mathbb{E}[e^{-\lambda X}\mid\mathcal{F}_t] =
\mathbb{E}[\mathbb{E}[e^{-\lambda X}\mid\mathcal{F}_{t+1}]\mid\mathcal{F}_t]$ by the tower property,
and the inner expectation is $e^{-\lambda\,\mathcal{U}_\lambda(X\mid\mathcal{F}_{t+1})}$. For the
second, a conditional $J_w$ is law invariant, cash invariant and coherent, and relevant, since $J_w(-\varepsilon\mathbf{1}_A) = -\varepsilon\beta(\mathbb{P}(A)) < 0$ whenever $\mathbb{P}(A) > 0$; and
\citet[Theorem~1.10]{kupper2009-law-invariant-time-consistent} show that a law-invariant, time-consistent
dynamic risk measure of this kind is entropic, and coherent only in the two limits $\lambda = 0$
and $\lambda = \infty$, the mean and the essential infimum. The latter has no weight satisfying
\eqref{eq:coherence-conditions}. The theorem is stated on a standard filtered probability space, hence
the qualifier \emph{on every filtration}.
\end{proof}

The failure has a concrete form in hedging. Let the hedging error be the sum of $N$ increments
$\xi_i = \sigma N^{-1/2}\zeta_i$, one per rebalancing date, with the $\zeta_i$ independent and
identically distributed with mean zero and variance one, and take expected shortfall at
$\alpha < 1$. Applied recursively, cash invariance, independence and positive homogeneity give
$\sum_{i=1}^N \mathcal{U}_\alpha(\xi_i) = \sqrt{N}\,\sigma\,\mathcal{U}_\alpha(\zeta_1)$, which diverges
to $-\infty$ as $N \to \infty$ since $\mathcal{U}_\alpha(\zeta_1) < 0$. Applied at inception to the total,
it converges to $-\sigma\,\varphi(\Phi^{-1}(\alpha))/\alpha$, the expected shortfall of a centred normal law with variance $\sigma^2$, where $\varphi$ is the standard normal density, since by the central limit theorem with unit variance the normalised sum converges in $\mathcal{W}_1$ and expected shortfall is $\mathcal{W}_1$-continuous (Proposition~\ref{prop:w1}). A
recursively applied preference therefore depends on how often the hedge is rebalanced, which is why
Problem~\ref{prob:ddh} is posed on the static objective.

A static objective still admits a recursion, on a larger state. The table records what each
preference requires; it is the organisation of the algorithms of Section~\ref{sec:algorithms}.
\begin{center}
\small
\begin{tabular}{llll}
\toprule
preference & time consistent & Bellman state & reference \\
\midrule
entropic & yes & $(X_t, \Delta_t)$ & cash invariance and (E3) \\
shortfall mixture, fixed thresholds & no & $(X_t, \Delta_t, W_t)$ & \citet{bauerle2011-mdp-average-value-at-risk}, one level \\
spectral, threshold function & no & $(X_t, \Delta_t, W_t)$ & \citet{bauerle2021-spectral-risk-measures-mdp} \\
spectral, applied recursively & yes & $(X_t, \Delta_t)$ & a different objective \\
\bottomrule
\end{tabular}
\end{center}

For the entropic utility the accumulated profit and loss drops out of the recursion, since
cash invariance moves $W_t$ outside the certainty equivalent. For expected shortfall and its
mixtures the Rockafellar--Uryasev representation turns the objective at fixed thresholds into the
expectation of a function of the terminal wealth, which has a Bellman equation once $W_t$ is part of
the state, and the thresholds are optimised outside the recursion; for a continuous spectrum they become a
function of the level \citep{bauerle2021-spectral-risk-measures-mdp}. The recursion is exact, but the
hedge it returns is not time consistent in the market-and-inventory state: re-solving at a later date with the
preference applied to the remaining profit and loss gives a different hedge, and $W_t$ is what
carries the commitment made at inception.

Exchanging the two suprema in \eqref{eq:es} makes this explicit for a single level:
\[
\sup_\pi \mathcal{U}_\alpha(Z^\pi) \;=\; \sup_{b \in \mathbb{R}}
\Big\{ b - \tfrac{1}{\alpha}\,\mathbb{E}_{S_0 \sim \rho_0}\big[V^b_0(S_0)\big] \Big\},
\qquad
V^b_t(s) \;=\; \inf_{a \in \mathcal{A}} \mathbb{E}_{s,a}\big[V^b_{t+1}(S_{t+1})\big],
\quad
V^b_T(s) \;=\; \big(b - v - g(s)\big)^+,
\]
where $v$ is the profit-and-loss coordinate of $s$, by \eqref{eq:pnl-is-W}. For fixed $b$ the inner
problem is an expected-utility problem on the state $(X_t, \Delta_t, W_t)$, and $b$ is optimised
outside it.

Two consequences follow for learning. First, a policy that improves a spectral objective at each
state separately, reading the critic's law at that state and weighting it by $w$, optimises neither
the static objective nor, in general, any stated one; it is valid only for the frozen expected-utility problem on the enlarged state. Where a functional is not
indifferent to mixtures there is a decision process in which a policy greedy with respect to the
optimal return distribution is not optimal \citep{avilapires2025-distributional-dynamic-programming}.
Second, the methods that differentiate the static objective as a whole need neither property, and
serve every preference; Proposition~\ref{prop:minorant} makes this precise. The algorithms of
Section~\ref{sec:algorithms} are grouped by these facts: those that differentiate the static
objective, those that solve its frozen expected-utility form on the enlarged state, and those that
use the entropic recursion on the market-and-inventory state.

\subsection{Estimability of a Preference}
\label{subsec:estimability}

The conditions of Section~\ref{subsec:admissible} are those of the classical theory, which applies a
risk measure to a law that is given. Here the law is supplied by a critic and is therefore
approximate, and errors in an estimated law are measured in the Wasserstein distance
$\mathcal{W}_1(X,Y) = \int_0^1 |F^{-1}_X - F^{-1}_Y|$. The objective must be continuous in that distance,
since otherwise a bound on the critic's error implies no bound on the error in the score.

\begin{proposition}[Continuity, and when it fails]
\label{prop:w1}
Let $w$ satisfy \eqref{eq:coherence-conditions}. If $\lVert w \rVert_\infty < \infty$ then
$|J_w(X) - J_w(Y)| \leq \lVert w \rVert_\infty \mathcal{W}_1(X,Y)$ for all $X,Y$ with finite first moment,
and no smaller constant works. If $\lVert w \rVert_\infty = \infty$ then $J_w$ is not
$\mathcal{W}_1$-continuous.
\end{proposition}

\begin{proof}
The bound is H\"older with exponents $(\infty,1)$. For sharpness let
$A = \{u : w(u) > \lVert w \rVert_\infty - \eta\}$, an initial interval since $w$ is
non-increasing, and take $F^{-1}_Y = F^{-1}_X - c\,\mathbf{1}_A$: then $\mathcal{W}_1 = c|A|$ while
$|J_w(X)-J_w(Y)| \geq c|A|(\lVert w \rVert_\infty - \eta)$. For the failure take $F^{-1}_{Y_n} = F^{-1}_X - M_n\mathbf{1}\{u<\varepsilon_n\}$ with
$M_n = 1/\beta(\varepsilon_n)$ and $\varepsilon_n \downarrow 0$: the gap is $M_n\beta(\varepsilon_n) = 1$, while
$\mathcal{W}_1 = \varepsilon_n/\beta(\varepsilon_n) \to 0$, because $\beta(\varepsilon)/\varepsilon \geq w(\varepsilon) \to \infty$.
\end{proof}

Boundedness is therefore a fourth requirement, independent of
\eqref{eq:coherence-conditions}: the power and Wang distortions of Section~\ref{subsec:ddh} are
coherent and fail it. The constant grows with the depth of the tail: for expected shortfall
$\lVert w \rVert_\infty = \alpha^{-1}$, so the bound on the error in the score varies inversely with the tail fraction.

A second consequence of learning the law constrains the \emph{order} of operations. A functional
is \emph{elicitable} when some loss is minimised in expectation exactly at its value, which is what
allows direct regression. \citet{ziegel2016-coherence-and-elicitability} proves that
among spectral functionals only the expected value is elicitable, so for every other $w$ the
objective $J_w$ is the target of no loss function. Quantiles are elicitable, so
$Z^\pi_\tau$ can be regressed on. Joint scores exist for finitely many levels, on laws with unique quantiles, when the matching quantiles are elicited alongside
\citep{fissler2016-higher-order-elicitability-osband}, but the levels must be fixed in advance.
Learning the law first and applying the weight afterwards is therefore the order that serves every
$w$ at once.

The law of the profit and loss is known only through simulated paths and a fitted critic, and a
preference evaluates that approximation through a weight on outcomes: the spectrum at the outcome's
rank, or its exponential tilt,
\begin{equation}
\label{eq:sample-weights}
\omega(z) \;=\; w\big(F_Z(z)\big)
\qquad\text{or}\qquad
\omega(z) \;=\; \frac{e^{-\lambda z}}{\mathbb{E}\big[e^{-\lambda Z}\big]} ,
\end{equation}
both non-negative and of mean one when $F_Z$ is continuous; each is the marginal utility of the
expected utility that bounds the preference from below (Proposition~\ref{prop:minorant}). Proposition~\ref{prop:w1} measured
the cost of a preference through $\lVert w \rVert_\infty$, the worst case over errors in the law.
The quantity that governs the typical case, both the number of paths a batch must contain and the
accuracy a critic must reach, is the second moment of the weight.

\begin{proposition}[Effective sample size]
\label{prop:ess}
Let $F_Z$ be continuous, let $G$ be a random vector with $\mathbb{E}[\lVert G\rVert^2 \mid Z] \leq
\kappa^2$, and let $(Z_j, G_j)_{j \leq B}$ be independent copies of $(Z,G)$. The weighted mean
$B^{-1}\sum_j \omega(Z_j)G_j$ has variance at most $\kappa^2\,\mathbb{E}[\omega(Z)^2]/B$, the bound for
an unweighted mean over $B/\mathbb{E}[\omega(Z)^2]$ paths, the effective sample size of importance
sampling \citep{kong1994-sequential-imputations,owen2013-monte-carlo}, and
\begin{enumerate}[label=(\roman*),topsep=2pt,itemsep=1pt,parsep=0pt,leftmargin=2em]
\item for a spectral preference, $\mathbb{E}[\omega(Z)^2] = \lVert w\rVert_2^2 = \iint
\max(\alpha,\alpha')^{-1}\,\mu(d\alpha)\,\mu(d\alpha')$, with $\mu$ as in Proposition~\ref{prop:mixture};
\item for the entropic utility, $\mathbb{E}[\omega(Z)^2] = \exp\big(2\lambda\,(\mathcal{U}_\lambda(Z) -
\mathcal{U}_{2\lambda}(Z))\big)$, which equals $e^{\lambda^2\sigma^2}$ when $Z$ is Gaussian with
variance $\sigma^2$.
\end{enumerate}
\end{proposition}
\begin{proof}
The variance is at most $B^{-1}\mathbb{E}[\omega^2\,\mathbb{E}[\lVert G\rVert^2 \mid Z]]$. (i)
$F_Z(Z)$ is uniform, so $\mathbb{E}[\omega^2] = \int_0^1 w^2$; substituting
$w(\tau) = \int_{(\tau,1]}\alpha^{-1}\mu(d\alpha)$ and integrating over $\tau < \min(\alpha,\alpha')$
gives $\iint \min(\alpha,\alpha')/(\alpha\alpha')\,\mu(d\alpha)\mu(d\alpha')$. (ii)
$\mathbb{E}[e^{-2\lambda Z}] = e^{-2\lambda\,\mathcal{U}_{2\lambda}(Z)}$ and
$\mathbb{E}[e^{-\lambda Z}]^2 = e^{-2\lambda\,\mathcal{U}_\lambda(Z)}$; for the Gaussian,
$\mathcal{U}_\lambda(Z) = \mathbb{E}[Z] - \lambda\sigma^2/2$.
\end{proof}

With $G$ the pathwise gradient of the profit and loss, the proposition bounds the variance of the
policy gradient under any preference. The same number bounds the critic's influence: pairing in
$L^2$ gives $|J_w(X) - J_w(Y)| \leq \lVert w\rVert_2\,\mathcal{W}_2(X,Y)$, and by
Proposition~\ref{prop:entropic} a perturbation $h$ of the quantile function moves
$\mathcal{U}_\lambda$ to first order by $\int w^\star h \leq \lVert w^\star\rVert_2\lVert h\rVert_2$, with $\lVert w^\star\rVert_2^2 = \mathbb{E}[\omega^2]$. A preference with a large factor therefore also amplifies the critic's error, by the square root of the same factor, and it
sets the bias of the batch estimates as well: the estimate $-\lambda^{-1}\log$ of a batch mean of $e^{-\lambda Z}$
overestimates $\mathcal{U}_\lambda$, an optimistic bias, by approximately $(\mathbb{E}[\omega^2] - 1)/(2\lambda B)$. To
match the variance bound of a risk-neutral run at batch $B_0$, a preference needs
$B_0\,\mathbb{E}[\omega^2]$ paths, or market paths drawn from a measure tilted toward the scenarios
it weights, which by (E1) can be reweighted without changing the policy.

The spectral factor $\lVert w\rVert_2^2$ is the same for every policy; the entropic factor
depends on the law of $Z^\pi$.

\section{Algorithms}
\label{sec:algorithms}

Problem~\ref{prob:ddh}, or its entropic counterpart with $\mathcal{U}_\lambda$ in place of $J_w$, asks for a
policy that maximises a preference $\mathcal{J}$ chosen among those of Section~\ref{sec:formulation}. The difficulty is that $\mathcal{J}$ is not an expectation: it depends on the whole law of the profit and loss,
through its ranks or through a logarithm, so the methods of reinforcement learning do not apply to it
directly. One inequality removes that difficulty by replacing the preference with an expected utility at the
current policy, and seven algorithms use it, each suited to what is available. They are built from three
ingredients. The first is a simulator of the market under (E1): the market evolves without
reference to the trades, so a market path can be drawn before the agent acts and replayed under a second
policy. The second is the sample weight $\omega$ of \eqref{eq:sample-weights}, the weight the preference
places on each outcome of the current policy. The third is a distributional critic
$Z_\tau(\cdot\,;\phi)$, whose output at the level $\tau_i = (i-\tfrac12)/M$, $i = 1, \dots, M$, estimates
the $\tau_i$-quantile of the return to go from a state. The critic is fitted with the quantile loss
$\rho_\tau(u) = u(\tau - \mathbf{1}\{u<0\})$ or its Huber-smoothed form
\citep{dabney2018-quantile-regression}. That loss contains no preference, so one critic can be read under
every preference. One result organises the section.

\begin{proposition}[Expected-utility minorant]
\label{prop:minorant}
Fix a policy $\pi_0$ with profit and loss $Z_0 = Z^{\pi_0}$. For a spectral preference with measure
$\mu$ as in Proposition~\ref{prop:mixture}, let $b_\alpha = F^{-1}_{Z_0}(\alpha)$ and
$\Phi_0(z) = \int \big(b_\alpha - \alpha^{-1}(b_\alpha - z)^+\big)\,\mu(d\alpha)$, with
$b_1 - (b_1 - z)^+$ read as $\min(b_1, z)$, which is $z$ when $b_1 = +\infty$; for the entropic
utility, let $\Phi_0(z) = \mathcal{U}_\lambda(Z_0) + \lambda^{-1}\big(1 - e^{-\lambda z}/
\mathbb{E}[e^{-\lambda Z_0}]\big)$. Then $\Phi_0$ is concave and non-decreasing with
$\Phi_0' = \omega$ (as a right derivative) at the law of $Z_0$. Assume $J_w(\pi_0) > -\infty$, respectively
$\mathbb{E}[e^{-\lambda Z_0}] < \infty$. Then for every policy $\pi$, writing $\mathcal{J}(\pi)$ for $J_w(\pi)$ or
$\mathcal{U}_\lambda(Z^\pi)$,
\[
\mathcal{J}(\pi) \;\geq\; \mathbb{E}\big[\Phi_0(Z^\pi)\big], \qquad \text{with equality at } \pi = \pi_0 ,
\]
the right-hand side possibly $-\infty$.
\end{proposition}
\begin{proof}
Expected shortfall is $\mathcal{U}_\alpha(X) = \max_b\{b - \alpha^{-1}\mathbb{E}[(b - X)^+]\}$,
attained at $F^{-1}_X(\alpha)$ \citep{rockafellar2000-optimization-of-cvar,rockafellar2002-cvar-general-loss-distributions}; fix $b = b_\alpha$ and
integrate against $\mu$. Since $z < b_\alpha$ exactly when $\alpha > F_{Z_0}(z)$,
$\Phi_0'(z) = \int_{(F_{Z_0}(z),1]}\alpha^{-1}\mu(d\alpha) = w(F_{Z_0}(z))$. For the entropic utility,
concavity of the logarithm gives $\log x \leq \log x_0 + x/x_0 - 1$; take
$x = \mathbb{E}[e^{-\lambda Z^\pi}]$, $x_0 = \mathbb{E}[e^{-\lambda Z_0}]$ and multiply by
$-\lambda^{-1}$.
\end{proof}

The minorant touches the objective at $\pi_0$ and lies below it elsewhere, which has two consequences.
Where both sides are differentiable, their difference is minimal at $\pi_0$, so the gradient of the
preference is the gradient of a risk-neutral problem with terminal utility $\Phi_0$; for coherent
measures the dual representation gives the same identity, up to terms from the constraints of the risk
envelope \citep{tamar2015-policy-gradient-coherent-risk}. The preference enters only through $\Phi_0$. Moreover, a policy that raises $\mathbb{E}[\Phi_0(Z)]$ by $\delta$ raises $\mathcal{J}$ by at
least $\delta$, which is the minorise--maximise principle \citep{hunter2004-tutorial-mm-algorithms}. Any
risk-neutral method that improves the frozen problem therefore improves the preference. The weight has
mean one under the law of $Z_0$ (for a spectral preference, when $F_{Z_0}$ is continuous), so the gradient
keeps the scale of the risk-neutral problem and a risk-neutral step size is a natural start.

What the minorant does not give is a recursion on the market-and-inventory state, since $\Phi_0$ acts on
the terminal profit and loss. Such a recursion needs time consistency, which among these preferences only
the entropic family has, the mean being its $\lambda = 0$ member (Proposition~\ref{prop:time-consistency}). Every other static
objective needs the enlarged state. For the entropic utility $\Phi_0$ is an affine transform of the
exponential utility, and $W_t$ drops out by (E3).

The algorithms therefore fall into four groups, listed in Table~\ref{tab:algorithms}. The three methods of
Group~I differentiate the static objective and differ in what the simulator allows. The two methods of
Group~II solve the frozen problem of a mixture of expected shortfalls on the enlarged state. Group~III
solves the entropic recursion, and Group~IV plans at the time of trading with the critic of a prior
policy at its leaves; the table also lists what each requires.

\vspace*{\fill}
\begin{table}[H]
\centering
\caption{The algorithms of this section. Under the entropic utility $W_t$ is dropped from the state. The
exact programmes and the planner use a one-step model of the market in place of paths.}
\label{tab:algorithms}
\setlength{\tabcolsep}{1pt}
\begin{tabular*}{\textwidth}{@{\extracolsep{\fill}}lllllll@{}}
\toprule
& & simulator & path & & & \\
& algorithm & gradient & replayed & critic & state & preferences \\
\midrule
I & policy search (\ref{subsec:policy-search}) & yes & no & none & $X_t,\Delta_t,W_t$ & any \\
  & policy gradient with a same-path baseline (\ref{subsec:policy-gradient}) & no & yes & none & $X_t,\Delta_t,W_t$ & any \\
  & advantage actor--critic (\ref{subsec:ac-advantage}) & no & no & values & $X_t,\Delta_t,W_t$ & any \\
II & policy search with shortfall thresholds (\ref{subsec:shortfall}) & yes & no & none & $X_t,\Delta_t,W_t$ & shortfall mixtures \\
  & dynamic programming for expected shortfall (\ref{subsec:es-dp}) & no & no & none & $X_t,\Delta_t,W_t$ & shortfall mixtures \\
III & dynamic programming for the entropic utility (\ref{subsec:entropic-dp}) & no & no & none & $X_t,\Delta_t$ & entropic \\
IV & planning with one distributional critic (\ref{subsec:planning}) & no & no & at the leaves & $X_t,\Delta_t,W_t$ & shortfall, entropic \\
\bottomrule
\end{tabular*}
\end{table}

\subsection{Policy Search with Spectral Weights}
\label{subsec:policy-search}

Deep hedging \citep{buehler2019-deep-hedging} differentiates the objective through a differentiable
simulator, as one global quantity and without a value function. The same approach serves every spectral
preference and the entropic utility. It requires rewards Lipschitz in the trade and differentiable almost
everywhere, as with proportional costs.

\begin{algorithm}[H]
\caption{Policy search with spectral weights}
\label{alg:policy-search}
\begin{algorithmic}[1]
\State \textbf{Input:} spectrum $w$ or risk aversion $\lambda$; batch size $B$; step size $\eta$
\For{iteration $= 1, 2, \dots$}
  \State simulate $B$ paths under $\pi_\theta$ and record $Z^{[1]}, \dots, Z^{[B]}$ \Comment{graph retained}
  \State \label{line:sort} sort $Z^{(1)} \leq \dots \leq Z^{(B)}$; set $u_{(j)} \gets (j - \tfrac12)/B$ \Comment{spectral only}
  \State \label{line:sort-weights}
  $\widehat{J} \gets \tfrac{1}{B}\sum_{j} w(u_{(j)})\, Z^{(j)}$ \ (spectral) \ or \
  $\widehat{J} \gets -\tfrac{1}{\lambda}\log \tfrac{1}{B}\sum_{k} e^{-\lambda Z^{[k]}}$ \ (entropic)
  \State ascend $\widehat{J}$ by a gradient step of size $\eta$ in $\theta$ \Comment{backpropagation through the path}
\EndFor
\State \textbf{Output:} the policy $\pi_\theta$
\end{algorithmic}
\end{algorithm}

Under a spectral preference the pathwise gradient keeps the form it has in deep
hedging, with each path's gradient multiplied by the spectrum at the path's rank. The result follows from
Proposition~\ref{prop:minorant}: the gradient of the preference is that of the frozen expected utility,
which is differentiated path by path. It needs a bounded weight.

\begin{proposition}[Pathwise gradient]
\label{prop:policy-gradient}
Let $\lVert w\rVert_\infty < \infty$, let $\theta \mapsto Z^{\pi_\theta}$ be almost surely Lipschitz near
$\theta_0$ with an integrable constant and differentiable at $\theta_0$, let $F_{Z^{\pi_{\theta_0}}}$ be
continuous, and let $\theta \mapsto J_w(\pi_\theta)$ be differentiable at $\theta_0$. Then
\begin{equation}
\label{eq:policy-gradient}
\nabla_\theta\, J_w\big(\pi_\theta\big)
\;=\; \mathbb{E}\Big[\, w\big(F_{Z^{\pi_\theta}}(Z^{\pi_\theta})\big)\,
\nabla_\theta Z^{\pi_\theta} \Big]
\qquad\text{at } \theta = \theta_0 .
\end{equation}
\end{proposition}
\begin{proof}
By Proposition~\ref{prop:minorant}, $J_w(\pi_\theta) - \mathbb{E}[\Phi_0(Z^{\pi_\theta})]$ is
non-negative and vanishes at $\theta_0$, so its gradient vanishes there. $\Phi_0$ is Lipschitz with
constant $\lVert w\rVert_\infty$ and differentiable off a set that $Z^{\pi_{\theta_0}}$ charges with
probability zero, so dominated convergence gives
$\nabla_\theta\mathbb{E}[\Phi_0(Z^{\pi_\theta})] = \mathbb{E}[\Phi_0'(Z)\nabla_\theta Z]$, and
$\Phi_0' = w \circ F_{Z^{\pi_{\theta_0}}}$.
\end{proof}

The dependence of the distribution function on $\theta$ contributes no term, as in the pathwise estimator
for expected shortfall of \citet{hong2009-simulating-sensitivities-cvar}. For the entropic utility the
same argument gives $\nabla_\theta\mathcal{U}_\lambda = \mathbb{E}[\omega(Z)\nabla_\theta Z]$ with the tilt
of \eqref{eq:sample-weights}, the pathwise form of Proposition~\ref{prop:entropic}, estimated by
self-normalised weights on the batch.

Holding later trades fixed, one more unit bought at date $t$ changes the
profit and loss by $H_T(X_T) - H_t(X_t)$, less the marginal costs of buying and liquidating it. The update
therefore moves the hedge toward the trades that would have gained on the paths the preference weights.
Under expected shortfall at $\alpha$ these are the worst $\alpha B$ paths of the batch, the ones on which
the book lost most. Under the entropic utility every path moves the hedge, with a weight proportional to
$e^{-\lambda Z}$, so the worst paths count most. The weights are recomputed on every batch: the update is
the gradient of the preference at the current policy, and recomputing them is what keeps the iteration on
the preference rather than on a fixed utility. No exploration is needed, since the simulator supplies fresh paths at every step,
and the algorithm differs from risk-neutral deep hedging only in lines~\ref{line:sort}--\ref{line:sort-weights}.

On a sorted batch the spectral objective is estimated by the
rank-weighted mean
\begin{equation}
\label{eq:rank-mean}
\widehat{J} \;=\; \frac{1}{B}\sum_{j=1}^{B} w\big(u_{(j)}\big)\, Z^{(j)},
\qquad u_{(j)} = \frac{j - \tfrac12}{B},
\end{equation}
a weighted average of order statistics that is consistent, with a bias of order $1/B$ for a bounded
weight. An unbounded weight still gives a finite estimate, but its bias decays more slowly and its largest
weight grows with $B$. The entropic estimator needs
no sorting. It is biased upward by about $(\mathbb{E}[\omega^2] - 1)/(2\lambda B)$
(Section~\ref{subsec:estimability}), which limits the risk aversion that a batch of a given size can
support. Under the entropic utility the paths may also be drawn from a measure tilted toward adverse
scenarios and reweighted by the likelihood ratio $L_k$ of the market path, which by (E1) does not involve
the policy. The estimate $-\lambda^{-1}\log\big(\sum_k L_k e^{-\lambda Z^{[k]}}/\sum_k L_k\big)$ remains
consistent, and the paths that carry the weight become more frequent.

With whole lots the simulator rounds each target to
the lot, and the gradient is passed straight through the rounding; a fee per ticket contributes no
gradient at all. The profit and loss is then piecewise constant in the trade and its law has atoms, so
the update is no longer the gradient of the objective. The direction it follows can still improve the
hedge, as Section~\ref{subsec:horizon} shows, and Section~\ref{subsec:policy-gradient} removes the
requirement altogether by never differentiating the reward.

\subsection{Policy Gradient with a Same-Path Baseline}
\label{subsec:policy-gradient}

Whole lots, a fee per ticket and black-box pricers leave the profit and loss without a usable derivative
in the trade, and these are the frictions of a real hedging book. A stochastic policy restores a gradient
without differentiating the environment, and a simulator that can replay a market path supplies a strong
baseline. The method serves every preference.

\begin{algorithm}[H]
\caption{Policy gradient with a same-path baseline}
\label{alg:policy-gradient}
\begin{algorithmic}[1]
\State \textbf{Input:} spectrum $w$ or risk aversion $\lambda$; batch $B$; exploration $\sigma$; clip $\epsilon$
\For{iteration $= 1, 2, \dots$}
  \State draw $B$ market paths \Comment{fixed before the agent acts, by (E1)}
  \State on each, roll targets $\Delta_{t+1} \sim \mathcal{N}(\mu_\theta(S_t), \sigma^2)$, giving $Z$, and
  $\mu_\theta$, giving $\bar Z$ \Comment{same market path}
  \State freeze $\Phi_0$ of Proposition~\ref{prop:minorant} on the batch; $\widehat A_t \gets
  \Phi_0(Z) - \Phi_0(\bar Z)$, normalised
  \State ascend the clipped surrogate $\sum_t \min\big(r_t \widehat A_t, \mathrm{clip}(r_t, 1 \pm \epsilon)
  \widehat A_t\big)$ for a few epochs
\EndFor
\State \textbf{Output:} the mean policy $\mu_\theta$
\end{algorithmic}
\end{algorithm}

The policy draws the target inventory $\Delta_{t+1} = \Delta_t + A_t$ from a
Gaussian of standard deviation $\sigma$ around the network's output $\mu_\theta(S_t)$, bounded to the
admissible inventories; $r_t$ is the ratio of $\pi_\theta(A_t \mid S_t)$ to its value at the parameters
that rolled the batch. By Proposition~\ref{prop:minorant} and the likelihood-ratio identity
\citep{williams1992-reinforce}, for any baseline $b_t$ that is a function of the market path and of the
history before $A_t$, wherever $\mathcal{J}(\pi_\theta)$ is differentiable,
\begin{equation}
\label{eq:lr-gradient}
\nabla_\theta \mathcal{J} \;=\; \mathbb{E}\Big[\sum_{t<T}\big(\Phi_0(Z) - b_t\big)\,
\nabla_\theta \log \pi_\theta(A_t \mid S_t)\Big].
\end{equation}
For expected shortfall this is the gradient of \citet{tamar2014-cvar-sampling}; for a coherent measure on a
finite sample space it is, up to terms from constraints that depend on $\theta$, the gradient under the
saddle-point density of \citet{tamar2015-policy-gradient-coherent-risk}; for the entropic utility it is
that of \citet{nass2019-entropic-risk-policy-search}. The estimator is unbiased but noisy. The hedging
environment supplies a baseline evaluated on the same market path.

\begin{proposition}[Same-path baseline]
\label{prop:same-path}
Under (E1), let $\bar Z$ be the profit and loss of the mean policy on the market path of $Z$. Then
$b_t = \Phi_0(\bar Z)$ leaves \eqref{eq:lr-gradient} unbiased, and it reduces the variance of the
$t$-th term whenever $\mathrm{Cov}(\Phi_0(Z)\,G_t, \Phi_0(\bar Z)\,G_t) > \tfrac12\mathrm{Var}(\Phi_0(\bar Z)\,G_t)$,
with $G_t$ the score, componentwise.
\end{proposition}
\begin{proof}
By (E1) the market path $\xi$ is drawn before any action, and $\bar Z$ is a function of $\xi$ alone.
The trade $A_t$ is drawn from $\pi_\theta(\cdot \mid S_t)$ with noise independent of $\xi$, so
$\mathbb{E}[G_t \mid \xi, S_t] = 0$ and $\mathbb{E}[\Phi_0(\bar Z)G_t] = 0$. The variance condition
is that of a control variate.
\end{proof}
Normalising the advantages and reading the thresholds of $\Phi_0$ from the same batch introduce a
dependence between paths that is not analysed here; both effects vanish as $B$ grows.

The correlation in the proposition tends to one as the exploration shrinks, since the two hedges then
trade almost alike on a market path they share. What remains in $\Phi_0(Z) - \Phi_0(\bar Z)$ is how much
better, in units of the preference, the explored hedge did than the unperturbed one on the same scenario.
In a hedge most of the variance of the profit and loss comes from the market path rather than from the
exploration, and the same-path baseline removes most of that part; a baseline that depends on the state
alone cannot, since it does not see the path to maturity. Each market path is therefore rolled twice, with
the explored policy and with the mean policy.
The reward is never differentiated, so any friction and any action constraint act only through the
simulated cash flows. Whole lots are rounded by the simulator and the score is taken of the unrounded draw,
so the estimator is unbiased for the objective smoothed by the exploration noise. The clip bounds how far
the policy moves per batch \citep{schulman2017-proximal-policy-optimization}, which keeps the frozen
$\Phi_0$ close to the minorant of the next policy. Under the entropic utility the paths may be tilted as in
Section~\ref{subsec:policy-search}; the likelihood ratio of the market path then enters the normaliser of
$\Phi_0$, the normalisation of the advantages and the clipped surrogate, and the mean policy is rolled on
the same tilted paths.

The method has two weaknesses. Under expected shortfall at $\alpha$ the advantage vanishes unless one of
the two hedges reaches the tail, a fraction of order $\alpha$ of the paths, so the batch must grow as
$1/\alpha$ to keep the signal. And the score of a Gaussian draw is of order $1/\sigma$, while a profit and
loss rounded to lots does not change in proportion to the draw, so $\sigma$ is kept above a floor
and the policy is scored by its mean $\mu_\theta$ rather than by the exploring policy. The method also needs a
simulator that can replay a market path, which live trading does not offer; there the baseline is lost and
a critic must replace it.

\subsection{Advantage Actor--Critic with a Distributional Critic}
\label{subsec:ac-advantage}

Algorithm~\ref{alg:ac-advantage} uses each market path once, as in live trading, where the path a hedge
meets cannot be replayed and the same-path baseline of Section~\ref{subsec:policy-gradient} is therefore
unavailable. A distributional critic, fitted with no preference, takes the place of that baseline and
enters through its values only. Since the critic carries the whole law of the return to go, one critic
serves every preference, and a change of preference changes only $\Phi_0$ in \eqref{eq:weighted-value}.

\begin{algorithm}[H]
\caption{Advantage actor--critic with a distributional critic}
\label{alg:ac-advantage}
\begin{algorithmic}[1]
\State \textbf{Input:} as Algorithm~\ref{alg:policy-gradient}; levels $\tau_i = (i-\tfrac12)/M$; trace parameter $\kappa \in [0,1]$; batches kept $n_c$
\For{iteration $= 1, 2, \dots$}
  \State roll $B$ paths under $\pi_\theta$, each once; keep the last $n_c$ batches
  \State critic steps: quantile regression of the realised return to go on these batches \Comment{no preference}
  \State freeze $\Phi_0$; $v_t \gets \tfrac{1}{M}\sum_i \Phi_0\big(W_t + Z_{\tau_i}(S_t;\phi)\big)$, \ $v_T \gets \Phi_0(Z)$;
  \ $\delta_t \gets v_{t+1} - v_t$
  \State $\widehat{A}_t \gets \sum_{k \geq 0}\kappa^k \delta_{t+k}$, normalised; ascend the clipped surrogate of
  Algorithm~\ref{alg:policy-gradient}
\EndFor
\State \textbf{Output:} the mean policy and the critic
\end{algorithmic}
\end{algorithm}

A critic $Z_\tau(s;\phi)$ of the return to go gives the weighted value
\begin{equation}
\label{eq:weighted-value}
v_t \;=\; \int_0^1 \Phi_0\big(W_t + Z_\tau(S_t;\phi)\big)\,d\tau ,
\qquad v_T = \Phi_0(Z),
\end{equation}
an estimate of $\mathbb{E}[\Phi_0(Z) \mid S_t]$: $W_t$ is the profit and loss already realised and
$Z_\tau(S_t;\phi)$ the law of what remains. The preference enters only through $\Phi_0$.

\begin{proposition}[Weighted advantage]
\label{prop:weighted-advantage}
Hold $\Phi_0$ fixed and let $v_t = \mathbb{E}[\Phi_0(Z) \mid S_t]$ exactly. Then for every
$\kappa \in [0,1]$ the generalised advantage $\widehat A_t = \sum_{k \geq 0}\kappa^k\delta_{t+k}$
\citep{schulman2015-gae} satisfies
$\mathbb{E}[\widehat A_t\, G_t] = \mathbb{E}[(\Phi_0(Z) - v_t)\,G_t]$, with $G_t$ the score of $A_t$,
so, before normalisation, it estimates the gradient \eqref{eq:lr-gradient} without bias.
\end{proposition}
\begin{proof}
Under $\pi_\theta$ the process $(v_t)$ is a martingale, by the Markov property and the tower property.
For $k \geq 1$ the information at date $t+k$ contains $A_t$, and
$\mathbb{E}[\delta_{t+k} \mid S_{0:t+k}, A_{0:t+k-1}] = 0$, so those terms vanish against $G_t$. The
term $k = 0$ has $\mathbb{E}[\delta_t \mid S_t, A_t] = \mathbb{E}[\Phi_0(Z) \mid S_t, A_t] - v_t$.
\end{proof}

The proposition holds for an exact critic. A fitted critic changes the variance of the estimate when it
is close to exact and adds a bias when it is not, and the trace parameter $\kappa$ decides how much of
either the estimate carries.

Besides reducing variance, a critic distributes credit across dates. The same-path difference credits
every trade of a path with one terminal outcome, while a critic of the return to go credits each trade
with the change it brings to the expected outcome. The gain is largest at the early dates of a long-dated
hedge, where the market noise after the trade dominates its effect. A trade at an early date is followed by many explored trades it
does not make, and the same-path difference charges all their effects to it; the one-step difference
$\delta_t$ charges only the next step, together with the error of the critic.

The trace parameter sets the balance. At $\kappa = 1$ the advantage is the Monte Carlo difference
$\Phi_0(Z) - v_t$, unbiased whatever the critic, which then acts as a baseline only; replacing the critic
by $\Phi_0(\bar Z)$ recovers the same-path difference of Section~\ref{subsec:policy-gradient}. At
$\kappa < 1$ the later market noise is down-weighted and the critic's error enters instead, amplified by
about the square root of the second moment of the weight on the law of what remains, which is largest
in the states the preference weights (Section~\ref{subsec:estimability}). The effect of one trade arrives mostly through the next price
increment, so short traces isolate it. Only the critic's values are used, never its gradient in the
action, which is harder to learn: the effect of a trade is of the order of one price increment, while the
values it would be differentiated from vary with the whole remaining market noise.

The critic needs the realised profit and loss only if the policy does. By (E3) the return to go of a policy of
$(X_t, \Delta_t)$ depends on $(X_t, \Delta_t)$ alone, and $W_t$ enters only through
$\Phi_0(W_t + \cdot\,)$ in \eqref{eq:weighted-value}; a policy that uses $W_t$ needs it in the critic as
well. The critic is also a valuation: it returns the law of what remains of the hedge at every visited
state. Under the entropic utility the weighted value factorises as
$v_t = \mathrm{const} - (\lambda\Lambda_0)^{-1} e^{-\lambda W_t}\int_0^1 e^{-\lambda
Z_\tau(S_t;\phi)}\,d\tau$, with $\Lambda_0 = \mathbb{E}[e^{-\lambda Z_0}]$, which separates the realised
profit and loss from the law of what remains. The critic is fitted on the last $n_c$ batches, which come
from slightly older policies, and the clip keeps that drift small. Marking the inventory at the current price moves $\Delta_t^\top H_t(X_t)$ from the
critic's target to the realised term and leaves $v_t$ unchanged. On tilted paths its loss would carry
the likelihood ratios to fit the law under the original measure; the experiments do not use this.

\subsection{Policy Search with Shortfall Thresholds}
\label{subsec:shortfall}

Algorithm~\ref{alg:shortfall} serves the mixtures of expected shortfall at finitely many levels,
$\mu = \sum_{k=1}^K c_k\,\delta_{\alpha_k}$. This is the form in which risk limits are usually stated, as a
regulatory charge at one level or an internal limit that combines several. Like the policy search of
Section~\ref{subsec:policy-search}, it needs a differentiable simulator.

\begin{algorithm}[H]
\caption{Policy search with shortfall thresholds}
\label{alg:shortfall}
\begin{algorithmic}[1]
\State \textbf{Input:} levels $\alpha_1 < \dots < \alpha_K$ with masses $c_k$; batch size $B$; step size $\eta$; period $m$
\For{iteration $n = 0, 1, 2, \dots$}
  \State simulate $B$ paths under $\pi_\theta$ and record $Z^{[1]}, \dots, Z^{[B]}$ \Comment{graph retained}
  \If{$m$ divides $n$}
    \State $b_k \gets$ the $\alpha_k$-quantile of $Z^{[1]}, \dots, Z^{[B]}$, for every $k$ \Comment{the full step}
  \EndIf
  \State ascend $\tfrac{1}{B}\sum_{j} u_b\big(Z^{[j]}\big)$ by a gradient step of size $\eta$ in $\theta$ \Comment{thresholds held fixed}
\EndFor
\State \textbf{Output:} the policy $\pi_\theta$ and the thresholds $b_k$
\end{algorithmic}
\end{algorithm}

For such a preference the minorant of Proposition~\ref{prop:minorant} is the
utility
\begin{equation}
\label{eq:shortfall-reduction}
u_b(z) \;=\; \sum_{k=1}^{K} c_k \Big( b_k - \frac{(b_k - z)^+}{\alpha_k} \Big),
\end{equation}
taken at thresholds $b = (b_1, \dots, b_K)$ equal to the quantiles of the current profit and loss, the term
at $\alpha_K = 1$ read as in Proposition~\ref{prop:minorant}. It is concave and piecewise linear, with
slope $\sum_k c_k\alpha_k^{-1}\mathbf{1}\{z < b_k\}$: the step weight of Section~\ref{subsec:admissible},
evaluated at the thresholds rather than at each path's rank.

\begin{proposition}[Reduction and ascent]
\label{prop:reduction}
For every policy, $J_w(\pi) = \max_{b}\mathbb{E}[u_b(Z^\pi)]$, attained at the quantiles
$b_k = F^{-1}_{Z^\pi}(\alpha_k)$, so $\sup_\pi J_w(\pi) = \sup_{\pi, b}\mathbb{E}[u_b(Z^\pi)]$. If $b$
are the quantiles of $Z^{\pi_n}$ and $\mathbb{E}[u_b(Z^{\pi_{n+1}})] \geq \mathbb{E}[u_b(Z^{\pi_n})]$,
then $J_w(\pi_{n+1}) \geq J_w(\pi_n)$. Where both are differentiable in the policy parameters, the
gradient of $J_w$ at $\pi_n$ is that of $\mathbb{E}[u_b(Z)]$ at these thresholds.
\end{proposition}
\begin{proof}
The first statement is Proposition~\ref{prop:minorant} with $\mu$ atomic, maximised over $b$
\citep{rockafellar2000-optimization-of-cvar}. With $b$ at the quantiles of $Z^{\pi_n}$,
$J_w(\pi_{n+1}) \geq \mathbb{E}[u_b(Z^{\pi_{n+1}})] \geq \mathbb{E}[u_b(Z^{\pi_n})] = J_w(\pi_n)$.
The last statement is the gradient consequence of Proposition~\ref{prop:minorant}.
\end{proof}

The algorithm is a block-coordinate ascent. Deep hedging optimises expected shortfall through the same
reduction, with simultaneous gradient steps in the policy and the threshold
\citep{buehler2019-deep-hedging}. Algorithm~\ref{alg:shortfall} replaces the step in $b$ by its exact
maximiser, the batch quantile, and then holds $b$ fixed for $m$ steps of
Algorithm~\ref{alg:policy-search}. The full step in $b$ is what the argument needs: at thresholds that
are not quantiles of the current law, the frozen objective lies strictly below $J_w$ and an improvement
of it says nothing about the preference. When every block improves the frozen objective, $J_w$ increases
at every refresh, and at a fixed point the policy is stationary for $J_w$.

Between two refreshes the problem is an expectation. The frozen utility reads as a mandate: every unit
lost below $b_k$ is charged at the rate $c_k/\alpha_k$, no gain above $b_K$ is credited, and a path below
$b_1$ is weighted by $\lVert w\rVert_\infty = \sum_k c_k/\alpha_k$. At a single level $\alpha < 1$ the
frozen objective is $b - \alpha^{-1}\mathbb{E}[(b - Z)^+]$, so the hedger maximises the threshold net of the
expected payoff of a put written on its own profit and loss and struck at it, scaled by $\alpha^{-1}$; several levels
add several such puts. The gradient weights each path by
$u_b'(Z)$, fixed for the whole block, so it is unbiased for the frozen objective and needs no sorting,
whereas the rank-weighted mean \eqref{eq:rank-mean} is biased at order $1/B$. Any risk-neutral method
solves the frozen problem, including the backward recursion of Section~\ref{subsec:es-dp} once $W_t$ is
in the state. The thresholds themselves come at no cost, since they are read from a sorted batch, and
they are the values at risk of the hedge at the charged levels.

The guarantee is that of an exact iteration, which the algorithm only approximates. A block of
stochastic steps need not improve the frozen objective, and the quantiles come from one batch, so the
policies need not converge. The block length $m$ sets a trade-off: a short block keeps the
thresholds close to the quantiles of the current policy, and a long block lets the policy improve the
frozen objective before the thresholds move. The optimal frozen hedge depends on $W_t$ (Section~\ref{subsec:es-dp}), and
the policy therefore needs it among its inputs. One threshold per level restricts the method to finitely
many levels; a continuous spectrum would require a threshold for every level.

\subsection{Dynamic Programming for Expected Shortfall}
\label{subsec:es-dp}

Algorithm~\ref{alg:dp} computes the references of Section~\ref{sec:experiments} for the mixtures of
expected shortfall at finitely many levels, exactly up to the quadrature and the interpolation. It needs
a one-step model of the market whose law can be integrated by quadrature (Gauss--Hermite in the
log-spot, mixed over jumps), a hedge with few rebalancing dates, and a state small enough for a grid.

\begin{algorithm}[H]
\caption{Dynamic programming for expected shortfall}
\label{alg:dp}
\begin{algorithmic}[1]
\State \textbf{Input:} level $\alpha$; initial state $(x_0, \Delta_0)$; a one-step model; grids on $x$ (log-spot), $\Delta$ and the cushion $e$
\State $\widetilde V_T(x, \Delta, e) \gets -\alpha^{-1}\big(e + c_T(x, -\Delta)\big)^+$ \Comment{terminal value}
\For{$t = T-1, \dots, 0$}
  \State $\bar V_t(x, \Delta', e) \gets \widehat{\mathbb{E}}\big[\widetilde V_{t+1}\big(X_{t+1}, \Delta', e - \Delta'(H_{t+1} - H_t) + C_{t+1} - C_t\big) \,\big|\, X_t = x\big]$
  \State $\widetilde V_t(x, \Delta, e) \gets \max_{\Delta'} \bar V_t\big(x, \Delta', e + c_t(x, \Delta' - \Delta)\big)$; \ $\pi_t$ the maximiser
\EndFor
\State \label{line:es-outer}
$e^\star \gets \argmax_e \big\{e + \widetilde V_0(x_0, \Delta_0, e)\big\}$; \ $b \gets e^\star + \Delta_0 H_0 - C_0$ \Comment{the threshold}
\State \textbf{Output:} the policies $\pi_t$, the threshold $b$ and the value $b + \widetilde V_0(x_0, \Delta_0, e^\star)$
\end{algorithmic}
\end{algorithm}

The frozen problem of Section~\ref{subsec:shortfall} acts on the terminal profit
and loss, so the state carries what has been accumulated. The recursion extends the single-level one of
Section~\ref{subsec:time-consistency} to mixtures, as an instance of
\citet{bauerle2011-mdp-average-value-at-risk} at one level and of
\citet{bauerle2021-spectral-risk-measures-mdp} in general.

\begin{proposition}[Augmented recursion]
\label{prop:es-dp}
Fix $b$. Let $V_T(x, \Delta, v) = u_b\big(v + g(x,\Delta)\big)$ and
\[
V_t(x, \Delta, v) \;=\; \sup_a\; \mathbb{E}\Big[V_{t+1}\big(X_{t+1},\, \Delta + a,\, v + R_t(x,a)\big)
\,\Big|\, X_t = x\Big].
\]
Then $\mathbb{E}[V_0(X_0, \Delta_0, 0)] = \sup_\pi \mathbb{E}[u_b(Z^\pi)]$ over all history-dependent
policies, attained by a Markov policy in $(X_t, \Delta_t, W_t)$ when the suprema are attained, as on a
finite set of targets, and $\sup_\pi J_w(\pi) = \sup_b \mathbb{E}[V_0(X_0, \Delta_0, 0)]$. For one level,
$V_t = b - \alpha^{-1}V^b_t$, with $V^b_t$ the recursion of Section~\ref{subsec:time-consistency}.
\end{proposition}
\begin{proof}
By \eqref{eq:pnl-is-W} and (E1)--(E3), $u_b(Z^\pi)$ is a function of the terminal state of a Markov
decision process on $(X_t, \Delta_t, W_t)$, whose value is given by the backward recursion and
attained by a Markov policy \citep{puterman1994-markov-decision-processes}. The second identity
exchanges the two suprema of Proposition~\ref{prop:reduction}
\citep{bauerle2011-mdp-average-value-at-risk}. For one level $u_b(z) = b - \alpha^{-1}(b - z)^+$, and
a supremum of $-\alpha^{-1}(\cdot)$ is $-\alpha^{-1}$ times an infimum.
\end{proof}

The recursion is solved on the \emph{cushion}, the threshold less the value of the book marked to market,
with the claim at its Black--Scholes price $C(t, X_t)$:
\begin{equation}
\label{eq:cushion}
e_t \;=\; b - W_t - \Delta_t H_t(X_t) + C(t, X_t)
\end{equation}
For a book short one claim with no cash before maturity, $b - Z = e_T + c_T(X_T, -\Delta_T)$ by (E2), and a
trade $A_t$ followed by the market step gives $e_{t+1} = e_t + c_t(X_t, A_t) - \Delta_{t+1}\big(H_{t+1}(X_{t+1})
- H_t(X_t)\big) + C(t+1, X_{t+1}) - C(t, X_t)$. Neither involves $b$, so for one level
$\widetilde V_t = -\alpha^{-1}V^b_t$ and one backward pass serves every threshold. The outer supremum
becomes the scalar maximisation of line~\ref{line:es-outer}, whose maximiser gives $b$, an
$\alpha$-quantile of the optimal hedge's profit and loss.

In Algorithm~\ref{alg:dp}, $H_t$ and $C_t$ abbreviate $H_t(X_t)$ and $C(t,X_t)$, the expectation
$\widehat{\mathbb{E}}$ integrates the one-step law by quadrature, interpolating the next value linearly in
$x$ and $e$, and $\Delta'$ ranges over a grid of inventories, or over the lots when trades are whole.
Mixtures are handled in the same way. With thresholds $b_k = b + o_k$ at fixed offsets $o_k$, each level
contributes its own terminal term, the recursion again does not involve $b$, and the offsets are searched
outside it. Each offset searched costs one backward pass, so the cost of a mixture grows with
the number of offsets tried.

Through the cushion, which carries the commitment made at inception, the optimal hedge depends on the
profit and loss already made. Far below the thresholds $u_b$ is linear and the hedger is risk neutral;
far above them, only the paths that fall back below the thresholds count. The dependence carries the commitment, not time
consistency: with the cushion in the state, the hedge chosen at inception stays optimal for the problem
frozen at $b$, but not for the preference re-applied at a later date. The recursive shortfall of Section~\ref{subsec:time-consistency} carries no
cushion, optimises a different objective, and loses against the static one
(Section~\ref{subsec:exact}). The method is limited to small
states, since the grid has one axis per coordinate. Any price equal to the payoff at $T$ gives the same
recursion, and the Black--Scholes price is chosen because it keeps the cushion small enough for a bounded
grid, beyond whose edges the value is held constant, an approximation that matters only on the paths
that reach them.

\subsection{Dynamic Programming for the Entropic Utility}
\label{subsec:entropic-dp}

Algorithm~\ref{alg:entropic-dp} computes the references for the entropic utility, at one risk aversion
per pass and exactly up to the quadrature and the interpolation. It needs a one-step model of the market
and a grid on the market state and the inventory; the expectation and the grids are those of
Algorithm~\ref{alg:dp} without the cushion.

\begin{algorithm}[H]
\caption{Dynamic programming for the entropic utility}
\label{alg:entropic-dp}
\begin{algorithmic}[1]
\State \textbf{Input:} risk aversion $\lambda$; initial state $(x_0, \Delta_0)$; a one-step model; grids on $x$ (log-spot) and $\Delta$
\State $\chi_T(x, \Delta) \gets g(x, \Delta)$ \Comment{liquidation value}
\For{$t = T-1, \dots, 0$}
  \State $\bar\chi_t(x, \Delta') \gets -\tfrac{1}{\lambda}\log \widehat{\mathbb{E}}\big[e^{-\lambda\chi_{t+1}(X_{t+1}, \Delta')} \,\big|\, X_t = x\big]$ \Comment{largest exponent factored out}
  \State $\chi_t(x, \Delta) \gets \max_{\Delta'}\big\{R_t(x, \Delta' - \Delta) + \bar\chi_t(x, \Delta')\big\}$; \ $\pi_t$ the maximiser \Comment{cash flows; no $W_t$}
\EndFor
\State \textbf{Output:} the policies $\pi_t$ and the value $\chi_0(x_0, \Delta_0)$
\end{algorithmic}
\end{algorithm}

The entropic family is the only family of law-invariant, relevant and
time-consistent preferences that are cash invariant (Section~\ref{subsec:entropic}). Cash invariance,
which also makes an indifference price independent of the initial wealth \citep{buehler2019-deep-hedging},
removes $W_t$: the entropic value of a position is its realised profit and loss plus a function of the
market state and the inventory.

\begin{proposition}[Entropic recursion]
\label{prop:entropic-dp}
Let $\chi_T(x,\Delta) = g(x,\Delta)$ and
\begin{equation}
\label{eq:entropic-bellman}
\chi_t(x, \Delta) \;=\; \max_{a}\; \Big\{ R_t(x,a) \;-\; \frac{1}{\lambda}\log\,
\mathbb{E}\Big[ e^{-\lambda\,\chi_{t+1}(X_{t+1},\, \Delta + a)} \,\Big|\, X_t = x \Big]\Big\}.
\end{equation}
If the exponential moments are finite and the maxima attained, as on a grid, the optimal conditional
value at date $t$ is $W_t + \chi_t(X_t, \Delta_t)$, the maximisers form an optimal Markov policy that
does not depend on $W_t$, and $\sup_\pi \mathcal{U}_\lambda(Z^\pi) = \mathcal{U}_\lambda\big(\chi_0(X_0,
\Delta_0)\big)$.
\end{proposition}
\begin{proof}
$\mathcal{U}_\lambda(Z) = \mathcal{U}_\lambda\big(-\lambda^{-1}\log\mathbb{E}[e^{-\lambda Z} \mid
\mathcal{F}_t]\big)$ by the tower property (Proposition~\ref{prop:time-consistency}), and
$\mathcal{U}_\lambda(c + X \mid \mathcal{F}_t) = c + \mathcal{U}_\lambda(X \mid \mathcal{F}_t)$ for
$\mathcal{F}_t$-measurable $c$. By (E2) $R_t$ is fixed by $(x,a)$ and by (E3) $W_t$ is additive, so both leave the logarithm,
and induction from $T$ gives the optimal value $W_t + \chi_t(X_t, \Delta_t)$ at every date, attained by
the maximisers; $\psi_t = e^{-\lambda\chi_t}$ solves the multiplicative recursion of a risk-sensitive
decision process \citep{howard1972-risk-sensitive-mdp}.
\end{proof}

The policy-evaluation form of this risk-averse Bellman equation is the one on which
\citet{murray2022-deep-hedging-continuous-rl} fit an actor and a critic; Algorithm~\ref{alg:entropic-dp} solves it on a grid instead.
The function $\chi_t$ has a direct reading. It is the certainty equivalent, at date $t$, of the profit
and loss still to come from a book holding $\Delta$ at $x$ and hedged optimally thereafter, so the entropic value of
the whole position is the profit and loss already realised plus $\chi_t$. The recursion therefore values
the book at every node of the grid, as the critic of Section~\ref{subsec:ac-advantage} does at the
states it visits.

The optimal hedge has no memory. It needs only $(X_t, \Delta_t)$, so Table~\ref{tab:algorithms} drops
$W_t$, and an early loss does not change the trades that follow. This is the opposite of the shortfall
hedge of Section~\ref{subsec:es-dp}, whose trades depend on the cushion. A desk can run it from the current position and the market alone, without
the history of the book. The risk aversion spans the
range of attitudes: as $\lambda \downarrow 0$ the bracket of \eqref{eq:entropic-bellman} tends to the
risk-neutral Bellman operator of Section~\ref{subsec:mdp}, and as $\lambda \to \infty$ it tends to $R_t$
plus the worst next value.

Under convex costs every $\chi_t$ is concave in the inventory, since $g$ and $R_t$ are and
$-\lambda^{-1}\log\mathbb{E}[e^{-\lambda\chi}]$ preserves concavity. For one instrument under a
proportional cost the maximiser is then a band: no trade while the inventory lies in an interval
$[\underline{\Delta}_t(x), \overline{\Delta}_t(x)]$, and otherwise a trade to its nearer edge, the
no-transaction band of the exponential-utility hedge \citep{whalley1997-asymptotic-hedging-transaction-costs}. A fixed charge per ticket breaks the concavity,
and Algorithm~\ref{alg:entropic-dp} then searches every target. Under stochastic volatility the variance
becomes a further coordinate of the grid.

The same recursion prices the claim. The premium at which a seller is indifferent to writing it is
$\chi_0$ without the claim minus $\chi_0$ with it, both computed by Algorithm~\ref{alg:entropic-dp}.
The recursion also returns a policy, and simulating that policy gives the law of its profit and loss and
with it the hedge's score under any other preference. Its one limitation against the shortfall recursion
is the scope of a pass: one pass serves a single risk aversion, whereas one pass of the shortfall
recursion serves every threshold, so a range of risk aversions costs one pass each. Because a pass returns both a value and a policy,
it can be checked against itself: the value $\mathcal{U}_\lambda(\chi_0)$ and the Monte Carlo score of
the simulated policy should agree up to the quadrature, the interpolation and the sampling error.
Section~\ref{subsec:networks} applies this check to the references and reports where it holds and
where it fails.

\subsection{Planning with One Distributional Critic}
\label{subsec:planning}

Algorithm~\ref{alg:planning} lets the preference be chosen when the trade is made. It needs a one-step
model of the market at the time of trading, a prior policy of $(X_t, \Delta_t)$, and one distributional
critic of that prior, fitted with no preference, and it serves the mixtures of expected shortfall and the
entropic utility.

\begin{algorithm}[H]
\caption{Planning with one distributional critic}
\label{alg:planning}
\begin{algorithmic}[1]
\State \textbf{Input:} critic $Z_\tau(\cdot\,;\phi)$ of a prior policy $\pi_0$, fitted with no preference; candidate targets; a one-step model
\State \textbf{Chosen when trading:} levels $\alpha_k$ and masses $c_k$, with $b_k$ first set to the prior's $\alpha_k$-quantiles on training paths, or risk aversion $\lambda$
\For{each date $t$, at the realised state $(X_t, \Delta_t, W_t)$}
  \For{each candidate target $\Delta'$}
    \State $v' \gets W_t + R_t(X_t, \Delta' - \Delta_t)$; \ $z_i \gets Z_{\tau_i}(X_{t+1}, \Delta'; \phi)$, $i \leq M$ \Comment{$z_i = g(X_T, \Delta')$ if $t+1 = T$}
    \State $Q(\Delta') \gets \widehat{\mathbb{E}}\big[\tfrac1M\sum_i u_b(v' + z_i)\big]$ \ or \ $v' - \tfrac{1}{\lambda}\log\widehat{\mathbb{E}}\big[\tfrac1M\sum_i e^{-\lambda z_i}\big]$ \Comment{law collapsed once}
  \EndFor
  \State trade to the candidate of largest $Q$ \Comment{depth one}
\EndFor
\State \textbf{Between passes} (shortfall): $b_k \gets$ the $\alpha_k$-quantiles of the planned hedge's profit and loss on training paths
\end{algorithmic}
\end{algorithm}

A fee per ticket or a minimum lot makes the objective multimodal, and a gradient method may stall
\citep{maggiolo2025-deep-hedging-non-convexity-alphazero}. Tree searches for an expected utility search
over trades instead \citep{szehr2021-mcts-hedging,maggiolo2025-deep-hedging-non-convexity-alphazero}.
Algorithm~\ref{alg:planning} searches to depth one: each candidate target is valued by a one-step
expectation, by quadrature, of a leaf that is the critic's law of the prior's return to go.

\begin{proposition}[Planning with exact leaves]
\label{prop:leaf}
Let $\pi_0$ be a policy of $(X_t, \Delta_t)$ whose targets are candidates, and let the critic be exact
for it, $Z_\tau(x, \Delta;\phi)$ being the $\tau$-quantile of the return to go of $\pi_0$ from
$(X_{t+1}, \Delta_{t+1}) = (x, \Delta)$, with the mean over levels replaced by the integral over $\tau$.
Let $\pi_1$ trade at every date to the candidate of largest $Q$, the one-step expectation being
exact. At fixed thresholds $b$, $\mathbb{E}[u_b(Z^{\pi_1})] \geq \mathbb{E}[u_b(Z^{\pi_0})]$, so
$J_w(\pi_1) \geq J_w(\pi_0)$ when $b$ are the quantiles of $Z^{\pi_0}$; under the entropic utility,
$\mathcal{U}_\lambda(Z^{\pi_1}) \geq \mathcal{U}_\lambda(Z^{\pi_0})$.
\end{proposition}
\begin{proof}
Since $\tau \sim U(0,1)$ reproduces the law, $\int_0^1 u_b(v + Z_\tau(x,\Delta;\phi))\,d\tau$ is the
value of $\pi_0$ at $(x, \Delta, v)$ in the recursion of Proposition~\ref{prop:es-dp}, and
$v - \lambda^{-1}\log\int_0^1 e^{-\lambda Z_\tau(x,\Delta;\phi)}\,d\tau$ its value in that of
Proposition~\ref{prop:entropic-dp}, so $Q$ is the action value of $\pi_0$ and $\pi_1$ is greedy at every date with respect to the value of
$\pi_0$, and backward induction, with the monotonicity of the conditional expectation, respectively of
$-\lambda^{-1}\log\mathbb{E}[e^{-\lambda\,\cdot}]$, makes the value of $\pi_1$ at least that of $\pi_0$
at every date \citep{puterman1994-markov-decision-processes}. Then
$J_w(\pi_1) \geq \mathbb{E}[u_b(Z^{\pi_1})] \geq \mathbb{E}[u_b(Z^{\pi_0})] = J_w(\pi_0)$ by
Proposition~\ref{prop:reduction}.
\end{proof}

The preference enters once, at the leaves, through fixed thresholds or the certainty equivalent. The
critic carries no preference, so the preference can be chosen at the time of trading without refitting
the critic. Applying $J_w$ afresh at every date would instead be the per-state rule of
Section~\ref{subsec:time-consistency}, which is not the static preference. The search compares trades on
scenarios and never differentiates the critic. With whole lots and a fee the candidates are the
admissible lots, no trade among them, and the search weighs the ticket paid now against the drift of an
unrebalanced hedge.

The planner and the actor--critic of Section~\ref{subsec:ac-advantage} use the critic in different ways.
The actor--critic uses its values to reduce the variance of a gradient and still learns a policy; the
planner uses them to choose among trades directly and learns none. Under the entropic utility, where the
planned hedge needs no $W_t$, it can serve as the next prior: refitting the critic to it and planning
again gives hedges each at least as good as the last when the critic is exact, a policy iteration. Whether
the sequence still improves with an approximate critic is open (Section~\ref{sec:discussion}).

The prior may be a delta hedge rounded to the lot. Its critic is fitted from states at random dates and
inventories, so that it is accurate where the candidates lead, and by (E3) it takes $(X_{t+1}, \Delta')$
alone. The one-step expectation is computed by quadrature rather than by sampling, because the largest of
several noisy estimates is biased upward and sampled scenarios would favour the noisiest candidates.
Quadrature removes that error and leaves only the critic's.

The thresholds start at the prior's quantiles, where the proposition applies, and move between passes to
those of the planned hedge. The leaves remain the prior's, so at the moved thresholds the proposition
still gives $\mathbb{E}[u_b(Z^{\pi_1})] \geq \mathbb{E}[u_b(Z^{\pi_0})]$, but no ascent of $J_w$.
Beyond one date the planned hedge is only as good as its leaves. The method is also the most expensive
per decision, since every candidate needs one evaluation of the critic at each quadrature node, and it is
the only one that needs the one-step model when trading. No bound is given for an approximate critic;
Section~\ref{subsec:onecritic} measures how the planner fares with a fitted one.

\section{Numerical Experiments}
\label{sec:experiments}

\subsection{Problem Setup}
\label{subsec:setup}

The theory of Sections~\ref{sec:formulation} and~\ref{sec:algorithms} makes predictions that can only be
checked on a problem whose optimum is known: that the preference changes the optimal hedge, that the
algorithms find that hedge, that each preference selects its own, and that one critic fitted without a
preference can serve several. We test them on the hedging problem of \citet{buehler2019-deep-hedging},
which is small enough for the programmes of Sections~\ref{subsec:es-dp} and~\ref{subsec:entropic-dp} to be
solved on a grid. Every learned hedge can then be compared with a reference, and not only with the delta
hedge, and the comparison separates what the algorithms get wrong from what the problem makes hard.
Section~\ref{subsec:networks} explains how the references are obtained and how far each can be trusted, and Table~\ref{tab:glance} summarises the findings and where each is shown.

\begin{table}[H]
\centering
\setlength{\tabcolsep}{4pt}
\caption{Results at a glance.}
\label{tab:glance}
\begin{tabularx}{\textwidth}{@{}LP{2.9cm}l@{}}
\toprule
finding & evidence & section \\
\midrule
The optimal hedge depends on the preference and on the past profit and loss; recursive shortfall loses $8.4\%$ & Figure~\ref{fig:structure} & \ref{subsec:exact} \\
Policy search is within $2\%$ of the reference in all twelve ten-date cells at $2{,}048$ steps & Tables~\ref{tab:campaign}, \ref{tab:converged} & \ref{subsec:learned} \\
Each preference selects its own hedge in all twelve pairwise comparisons & Figure~\ref{fig:preference}, Table~\ref{tab:cross} & \ref{subsec:distribution} \\
At thirty dates policy search stays within $2.8\%$; with lots and a fee the likelihood-ratio methods lead & Tables~\ref{tab:T30}, \ref{tab:lots} & \ref{subsec:horizon} \\
One critic fitted without a preference reads nine preferences and supports planning & Figure~\ref{fig:critic}, Table~\ref{tab:planning} & \ref{subsec:onecritic} \\
Gradient variance tracks $\mathbb{E}[\omega^2]$, and tilted paths lower it & Figures~\ref{fig:cost}, \ref{fig:history} & \ref{subsec:cost} \\
\bottomrule
\end{tabularx}
\end{table}

The agent is short one European call at the money, $K = S_0$, with unit notional, a maturity of thirty
days and payoff $(S_T - K)^+$. It hedges by trading the underlying at $T$ equally spaced dates, ten or
thirty. The same claim at the same maturity is the example of \citet{buehler2019-deep-hedging}. Its delta hedge
is known in closed form and gives a baseline, and its payoff is convex, so the losses of a poorly hedged
book gather in one tail, where the preferences differ most. The premium is left out, since by cash invariance it shifts every score equally. The environment
of Section~\ref{subsec:trading-mdp} is specialised as follows, with $S_t$ denoting the spot price in this
section. The book generates no cash before maturity, so $D_t = 0$ for $t < T$ and $D_T = -(S_T - K)^+$.
The hedging instrument is the underlying, $H_t(X_t) = S_t$ with $d_h = 1$, and from $\Delta_0 = 0$ the
target inventory lies in $[-0.1, 1.1]$, in multiples of a lot size $\ell$ in the lots problem. The friction
is proportional at a rate $\gamma$, with a fee $f$ per ticket in one problem, so that
$c_t = \gamma|A_t|S_t + f\,\mathbf{1}\{A_t \neq 0\}$ for $t < T$; the liquidation at maturity pays the
proportional cost alone.

Three price models share $S_0 = K = 1$, zero drift, a maturity of $30/365$ years and a total annual
variance of $0.04$, so they differ in the shape of the tail and not in its scale. The first is geometric
Brownian motion, $dS_t = \sigma S_t\,dB_t$ with $\sigma = 0.20$. The second is Heston,
$dS_t = \sqrt{v_t}\,S_t\,dB^S_t$ and $dv_t = k(\bar v - v_t)\,dt + \xi\sqrt{v_t}\,dB^v_t$, with
$v_0 = \bar v = 0.04$, $k = 2$, $\xi = 0.3$ and correlation $\varrho = -0.7$ between the drivers; the variance
is then a coordinate of $X_t$. The third is Merton, with a diffusive volatility $\sigma_D = 0.166$, jumps at
Poisson intensity $\nu = 1$, normal log-jumps of mean $m_J = -0.05$ and deviation $s_J = 0.10$, and a drift
that makes the spot a martingale. A jump
falls on about one path in thirteen and thickens the left tail (Figure~\ref{fig:setting}a). Heston thus
adds a variance that rises as the spot falls, and Merton adds moves that no rebalancing can follow. The
base problem trades at ten or thirty dates, ten only on Merton, with a cost of $\gamma = 1\%$. The lots
problem, on Brownian motion at thirty dates and at ten for planning, has lots of $\ell = 0.1$, a cost of
$\gamma = 0.2\%$ and a fee of $f = 0.0005$ per ticket.

The preferences are expected shortfall at $50\%$ and $10\%$, the two-level mixture with half its weight
at $1\%$ and half at $10\%$, the Wang distortion at $a = 1$, and the entropic utility at $\lambda = 25$ and
$50$. They span the range of tail weights, from expected shortfall at $50\%$, which weights the worse half
of the paths equally, to the two-level mixture, whose weight on the worst $1\%$ is $55$, and the entropic
utility, which weights every path and the worst most. The baselines are the Black--Scholes delta hedge, rounded to the lot when trades are whole, and no
hedge. All hedges and references of a problem share $200{,}000$ test paths, on which the standard error
of a reference is $0.1$ to $0.3\%$ of its value on Brownian motion and $1.3\%$ on Merton at $\lambda = 50$. Scores are utilities of the profit and loss, so a higher score is better throughout. Every hedge is scored
under the preference it was trained for, except in Section~\ref{subsec:distribution}, where each is scored
under all of them. Each result reports the median over seeds of the score,
in percent of the spot, the gap to the reference, in percent of its magnitude, and the share of the delta
hedge's gap that the learned hedge closes. A hedge that closes the whole gap matches the reference, and one that
closes none does no better than the delta hedge. The Wang distortion has no reference and is reported by its
score alone.

\subsection{Networks and Training}
\label{subsec:networks}

The policy is a multilayer perceptron with two hidden layers of width $64$ and $\tanh$ activations, the
same for every model, preference and algorithm. Its output is a target inventory
bounded to the admissible interval. Its inputs are the log-moneyness, the time to maturity, the
inventory, the variance on Heston and, under a spectral preference, the profit and loss, on which the
shortfall hedge depends (Section~\ref{subsec:es-dp}). Under the entropic utility the profit and loss is
omitted, since the optimal hedge does not depend on it (Proposition~\ref{prop:entropic-dp}).

Training starts from random weights. Each step draws $4{,}096$ fresh paths and
applies the algorithm's update with Adam, at a learning rate decayed by a cosine schedule from
$3 \cdot 10^{-3}$ to $3 \cdot 10^{-4}$. Every algorithm of Group~I and policy search with thresholds is
trained for $512$ steps under every preference it serves; policy search and the policy gradient also for
$2{,}048$ steps, the budget at thirty dates, as is the actor--critic on the lots problem. Each configuration has three seeds and is reported at its last
step. The policy gradient and the actor--critic are scored by their mean policy $\mu_\theta$.

\begin{table}[H]
\centering
\caption{Settings particular to each algorithm.}
\label{tab:settings}
\begin{tabularx}{\textwidth}{@{}P{3.3cm}L@{}}
\toprule
algorithm & settings \\
\midrule
policy search & rank weights \eqref{eq:rank-mean} or batch certainty equivalent; a straight-through estimator for the rounding to lots \\
policy search with & refreshed to the batch quantiles every $m = 16$ steps \\
\quad shortfall thresholds & \\
policy gradient & exploration s.d.\ $0.05$, clip $\epsilon = 0.2$, $4$ epochs over minibatches of $8{,}192$ decisions \\
actor--critic & as above; critic with $M = 32$ levels, $20$ steps per batch on the last $n_c = 4$ batches, $\kappa = 0.9$ \\
planning & critic of the rounded delta hedge with $M = 32$ levels; $31$ quadrature nodes \\
\bottomrule
\end{tabularx}
\end{table}

The critic of the actor--critic is the same network with $M$ outputs, one per level
$\tau_i = (i - \frac12)/M$, fitted by the Huber-smoothed quantile loss to the returns to go of the last four
batches. Its loss contains no preference, which enters only when the frozen $\Phi_0$ of
Proposition~\ref{prop:minorant} is applied to each output in \eqref{eq:weighted-value}. Under expected
shortfall at $\alpha$ the value of a state is the batch quantile $b$ less $\alpha^{-1}$ times the average
of $(b - W_t - Z_{\tau_i})^+$, so only the levels below the threshold count. Under the entropic utility it
is affine in $e^{-\lambda W_t}$ times the average of $e^{-\lambda Z_{\tau_i}}$, so every level counts and the
worst counts most. The planner's critic is fitted in the same way to the rounded delta hedge, from random
dates and inventories.

The references come from the programmes of Sections~\ref{subsec:es-dp} and~\ref{subsec:entropic-dp},
solved on a grid; each reference is the Monte Carlo value of the grid policy on the test paths. On
Brownian motion at ten dates with a cost of $1\%$, under expected shortfall at $50\%$ and $10\%$, the
entropic utility at $\lambda = 10$ and $50$ and the two-level mixture, the grid value and the Monte Carlo
value agree within $1\%$, as they do on Merton at $\lambda = 25$: the reference is self-consistent to that
accuracy. Coarsening the inventory grid
to thirteen targets moves the grid value at $\lambda = 10$ by $0.03\%$ and its Monte Carlo value by $0.04\%$. Every other reference is the
value of a grid policy, unchecked or failing the check; learned hedges can beat such a reference.

\begin{figure}[H]
\centering
\includegraphics[width=\textwidth]{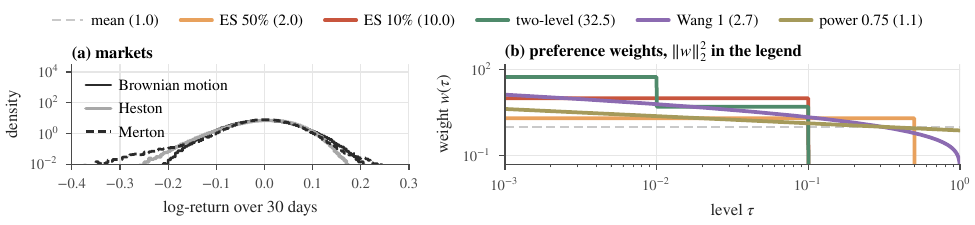}
\caption{(a) Densities of the log-return over thirty days. (b) Weights of the preferences, with
$\lVert w\rVert_2^2$ in the legend; power $0.75$ is only scored.}
\label{fig:setting}
\end{figure}

The weights of Figure~\ref{fig:setting}b also bound how much each preference costs to learn. By
Proposition~\ref{prop:ess} a weighted gradient on $4{,}096$ paths has the variance bound of an unweighted
mean over $4{,}096/\lVert w\rVert_2^2$ paths. That effective batch is $2{,}048$ paths under expected
shortfall at $50\%$, about $410$ at $10\%$, and about $126$ under the two-level mixture. The mixture therefore learns
from about sixteen times fewer effective paths than expected shortfall at $50\%$, and
Section~\ref{subsec:cost} measures what that costs.

\subsection{Hedges from Dynamic Programming}
\label{subsec:exact}

\begin{figure}[H]
\centering
\includegraphics[width=\textwidth]{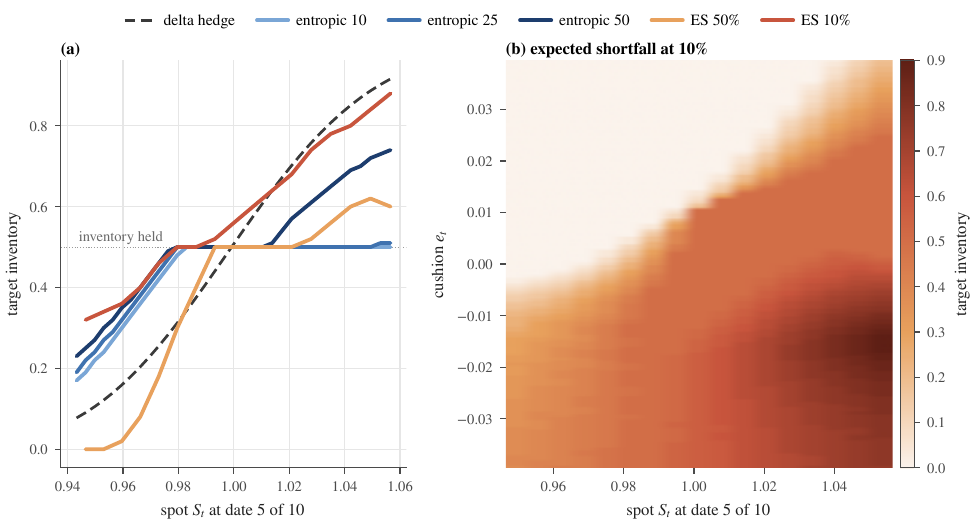}
\caption{Hedges of the grid programmes on Brownian motion at ten dates, cost $1\%$, at date $5$ with an
inventory of $0.5$ held before the trade. (a) Target inventory against the spot under each preference,
the shortfall hedges at the median cushion their own policy visits; the grid at $\lambda = 25$ is not
checked. (b) Target inventory under expected shortfall at $10\%$ against the spot and the cushion
\eqref{eq:cushion}.}
\label{fig:structure}
\end{figure}

Before any learning enters, does the preference change the optimal hedge? The grid programmes of
Sections~\ref{subsec:es-dp} and~\ref{subsec:entropic-dp} answer this directly on Brownian motion at ten
dates with a cost of $1\%$ (Figure~\ref{fig:structure}).

The references are accurate enough to compare against. The grid value and the Monte Carlo value of its
policy are $-4.163$ and $-4.197$ under expected shortfall at $50\%$, a discrepancy of $0.8\%$; they agree
to $0.04\%$ at $10\%$, and to $0.3\%$ and $0.01\%$ under the entropic utility at $\lambda = 10$ and $50$.
The tables report the Monte Carlo value, which is
the score of a policy that can actually be run.

The two families produce hedges of different shapes (panel a). The entropic hedges are bands: no trade
near the money, and otherwise a trade to the nearer edge, with a band that narrows as $\lambda$ grows
(Section~\ref{subsec:entropic-dp}). A narrower band tolerates less drift from the target before
the hedge pays to trade. Below the money they hold more than the delta; above it they barely
trade at $\lambda = 10$ and $25$, and at $\lambda = 50$ the hedge buys but stays below the delta. The
shortfall hedges are shaped otherwise. At $50\%$ the hedge sells its whole inventory where the spot is
lowest and stays below the delta above the money, while at $10\%$ it holds more than the delta below the
money and follows it above. Within each family, more weight on losses means more trading: over ten
dates the hedges trade $1.83$ times per path at $\lambda = 10$ and $4.65$ at $\lambda = 50$, against
$4.55$ under expected shortfall at $50\%$ and $7.12$ at $10\%$. Trading costs money, so the more
risk-averse hedges give up expected profit and loss to narrow the tail.

The shortfall hedge also depends on the past (panel b). Above the money its target runs from near zero,
where the cushion is large, to near $0.9$. Across the central $90\%$ of the cushions it visits, the range
of its target averages $0.16$ and is positive at $98\%$ of the states. No policy of $(X_t, \Delta_t)$
alone represents it (Proposition~\ref{prop:es-dp}), which is why learned policies receive the profit and
loss under a spectral preference.

Ignoring that dependence has a measurable cost. Applying expected shortfall at each date to the one-step
law gives a time-consistent hedge on $(X_t, \Delta_t)$, but for a different objective. That hedge treats
each date as if the losses already made did not matter. Scored under the
static objective it reaches $-4.550$ against the reference's $-4.197$ at $50\%$, and $-5.538$ against
$-5.108$ at $10\%$. At both levels it is $8.4\%$ below the reference, about half the distance to the delta hedge.
The preference therefore shapes the optimal hedge, and the static objective has to be solved on the
enlarged state or differentiated as a whole.

\subsection{Learned Hedges and the Reference}
\label{subsec:learned}

Do the learning algorithms of Section~\ref{sec:algorithms} find these hedges? Tables~\ref{tab:campaign}
and~\ref{tab:converged} score them on the three ten-date problems, after $512$ and $2{,}048$ steps.

\begin{table}[H]
\centering
\footnotesize
\caption{Median score and gap (\%) at $512$ steps, ten dates, cost $1\%$. $^\dagger$Value of a grid
policy failing the $1\%$ check; Table~\ref{tab:converged} gives the Heston shortfall gaps at $2{,}048$ steps,
Section~\ref{subsec:distribution} the two-level gaps.}
\label{tab:campaign}
\begin{tabular}{lrrrrrr}
\toprule
\textit{Brownian motion} & \makebox[43pt][r]{ES 50\%} & \makebox[43pt][r]{ES 10\%} & \makebox[43pt][r]{two-level} & \makebox[43pt][r]{Wang 1} & \makebox[43pt][r]{entropic 25} & \makebox[43pt][r]{entropic 50} \\
\midrule
policy search & -4.22 (0.6) & -5.25 (2.8) & -5.64 & -4.66 & -3.58 (1.3) & -3.87 (1.5) \\
policy gradient & -4.26 (1.6) & -5.22 (2.2) & -5.63 & -4.67 & -3.61 (2.1) & -3.88 (1.7) \\
actor--critic & -4.27 (1.8) & -5.21 (2.0) & -5.65 & -4.68 & -3.62 (2.5) & -3.89 (1.9) \\
thresholds & -4.22 (0.5) & -5.25 (2.7) & -5.63 & -- & -- & -- \\
\midrule
reference & -4.20 & -5.11 & -5.52 & -- & -3.53 & -3.81 \\
delta hedge & -4.96 & -5.98 & -6.59 & -5.22 & -4.35 & -4.47 \\
no hedge & -4.59 & -10.45 & -13.40 & -6.40 & -4.86 & -9.85 \\
\bottomrule
\end{tabular}
\\[4pt]
\begin{tabular}{lrrrrrr}
\toprule
\rlap{\textit{Merton}}\phantom{\textit{Brownian motion}} & \makebox[43pt][r]{ES 50\%} & \makebox[43pt][r]{ES 10\%} & \makebox[43pt][r]{two-level} & \makebox[43pt][r]{Wang 1} & \makebox[43pt][r]{entropic 25} & \makebox[43pt][r]{entropic 50} \\
\midrule
policy search & -4.05 (0.5) & -6.44 (2.8) & -8.95 & -5.15 & -3.67 (2.6) & -5.37 (0.7) \\
policy gradient & -4.12 (2.1) & -6.38 (1.9) & -8.93 & -5.17 & -3.72 (3.9) & -6.04 (13.3) \\
actor--critic & -4.10 (1.8) & -6.45 (3.0) & -8.99 & -5.18 & -3.75 (4.9) & -6.01 (12.8) \\
thresholds & -4.05 (0.5) & -6.44 (2.8) & -8.94 & -- & -- & -- \\
\midrule
reference & -4.03 & -6.26 & -8.82$^\dagger$ & -- & -3.58 & -5.33$^\dagger$ \\
delta hedge & -4.83 & -6.81 & -10.33 & -5.85 & -4.79 & -11.01 \\
no hedge & -4.28 & -9.66 & -13.35 & -6.14 & -7.95 & -25.94 \\
\bottomrule
\end{tabular}
\\[4pt]
\begin{tabular}{lrrrrrr}
\toprule
\rlap{\textit{Heston}}\phantom{\textit{Brownian motion}} & \makebox[43pt][r]{ES 50\%} & \makebox[43pt][r]{ES 10\%} & \makebox[43pt][r]{two-level} & \makebox[43pt][r]{Wang 1} & \makebox[43pt][r]{entropic 25} & \makebox[43pt][r]{entropic 50} \\
\midrule
policy search & -4.16 & -5.28 & -5.71 & -4.62 & -3.47 (1.8) & -3.79 (1.7) \\
policy gradient & -4.20 & -5.25 & -5.72 & -4.63 & -3.49 (2.4) & -3.80 (2.0) \\
actor--critic & -4.21 & -5.25 & -5.71 & -4.64 & -3.49 (2.2) & -3.80 (1.9) \\
thresholds & -4.16 & -5.28 & -5.71 & -- & -- & -- \\
\midrule
reference & -4.14$^\dagger$ & -5.22$^\dagger$ & -- & -- & -3.41 & -3.73 \\
delta hedge & -4.94 & -6.02 & -6.75 & -5.24 & -4.35 & -4.47 \\
no hedge & -4.54 & -9.45 & -11.62 & -5.91 & -4.07 & -6.67 \\
\bottomrule
\end{tabular}

\vspace{2pt}
\caption{As Table~\ref{tab:campaign}, at $2{,}048$ steps; a negative gap is a
score above the reference.}
\label{tab:converged}
\begin{tabular}{llrrrr}
\toprule
market & algorithm & ES 50\% & ES 10\% & entropic 25 & entropic 50 \\
\midrule
Brownian motion & policy search & -4.20 (0.0\%) & -5.17 (1.1\%) & -3.55 (0.5\%) & -3.84 (0.7\%) \\
Brownian motion & policy gradient & -4.25 (1.3\%) & -5.18 (1.4\%) & -3.61 (2.1\%) & -3.87 (1.6\%) \\
Merton & policy search & -4.04 (0.3\%) & -6.31 (0.8\%) & -3.65 (1.9\%) & -5.39 (1.1\%) \\
Merton & policy gradient & -4.11 (2.0\%) & -6.34 (1.3\%) & -3.70 (3.3\%) & -5.86 (10.0\%) \\
Heston & policy search & -4.14 (-0.1\%) & -5.20 (-0.4\%) & -3.44 (0.9\%) & -3.75 (0.5\%) \\
Heston & policy gradient & -4.19 (1.2\%) & -5.21 (-0.2\%) & -3.49 (2.3\%) & -3.79 (1.8\%) \\
\bottomrule
\end{tabular}

\end{table}

Every learned hedge beats the delta hedge on every seed, a weak test under expected shortfall at $50\%$,
where the delta hedge scores below no hedge in every market.

After $512$ steps, in the cells whose gap the table shows, policy search is $0.5$ to $2.8\%$ from the reference, with or
without shortfall thresholds, which agree within $0.01$ in every cell (Proposition~\ref{prop:reduction}).
The policy gradient is $1.6$ to $3.9\%$ from it and the actor--critic $1.8$ to $4.9\%$, except on Merton at
$\lambda = 50$, where both are $13\%$ off. After $2{,}048$ steps policy search is within $2\%$ in all
twelve cells and within $1\%$ in nine, its medians moving by at most $0.12\%$ with two more seeds, and the
policy gradient is $-0.2$ to $3.3\%$ from the reference, again except on Merton at $\lambda = 50$
($10.0\%$). On Heston under expected shortfall, whose reference fails the check, policy search scores
slightly above it.

Merton at $\lambda = 50$ is hard for the likelihood-ratio
methods but not for policy search. The jumps concentrate the entropic weight on a few tail paths, which
raises the variance of a likelihood-ratio gradient: the seeds spread over $0.40$ and $0.57$ at $512$
steps, against $0.04$ for policy search; tilted paths help both methods tested on them
(Section~\ref{subsec:cost}).

\subsection{Comparison across Preferences}
\label{subsec:distribution}

Does each preference produce its own learned hedge, and how does the weight shape the law of the profit
and loss? Policy search is trained once per preference and each hedge is scored under every preference
(Table~\ref{tab:cross}).

\begin{figure}[H]
\centering
\includegraphics[width=\textwidth]{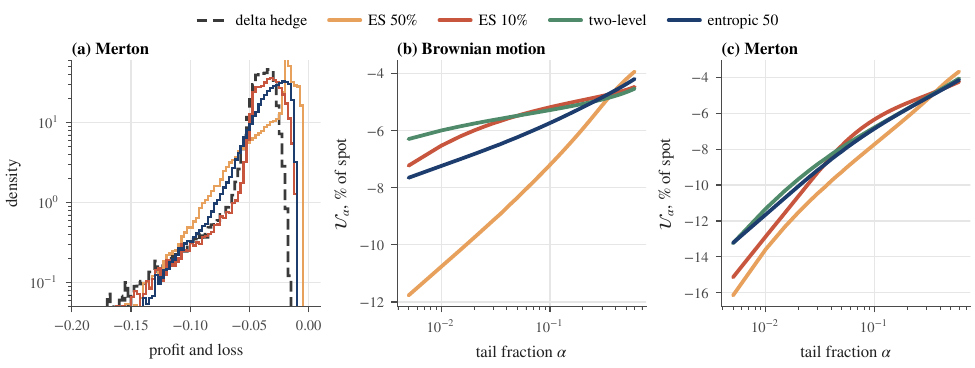}
\caption{Hedges learned by policy search at $2{,}048$ steps, ten dates, cost $1\%$, one per preference.
(a) Densities of the profit and loss on Merton, with that of the delta hedge. (b, c) Shortfall curves
$\alpha \mapsto \mathcal{U}_\alpha$ of the same hedges and of the two-level hedge, absent from (a), on
Brownian motion and on Merton.}
\label{fig:preference}
\end{figure}

\begin{table}[H]
\centering
\footnotesize
\setlength{\tabcolsep}{1.6pt}
\caption{Policy search at $512$ steps, trained under the preference of each row and scored under that
of each column: medians in units of $1\%$ of the spot. The best score in each column is in bold.}
\label{tab:cross}
\begin{tabular}{@{}c@{\hspace{0.8em}}c@{}}
\textit{Brownian motion} & \textit{Merton} \\
\begin{tabular}{lrrrr}
\toprule
trained for $\backslash$ scored under & ES 50\% & ES 10\% & two-level & Wang 1 \\
\midrule
ES 50\% & \textbf{-4.22} & -7.46 & -9.34 & -5.30 \\
ES 10\% & -4.57 & \textbf{-5.25} & -5.70 & -4.73 \\
two-level & -4.64 & -5.31 & \textbf{-5.64} & -4.75 \\
Wang 1 & -4.45 & -5.35 & -5.84 & \textbf{-4.66} \\
\bottomrule
\end{tabular}
 & \begin{tabular}{lrrrr}
\toprule
trained for $\backslash$ scored under & ES 50\% & ES 10\% & two-level & Wang 1 \\
\midrule
ES 50\% & \textbf{-4.05} & -7.72 & -10.49 & -5.41 \\
ES 10\% & -4.43 & \textbf{-6.44} & -9.50 & -5.37 \\
two-level & -4.32 & -6.79 & \textbf{-8.95} & -5.22 \\
Wang 1 & -4.16 & -6.81 & -9.06 & \textbf{-5.15} \\
\bottomrule
\end{tabular}
 \\ \noalign{\vskip 5pt}
\multicolumn{2}{c}{\textit{Heston}} \\
\multicolumn{2}{c}{\begin{tabular}{lrrrr}
\toprule
trained for $\backslash$ scored under & ES 50\% & ES 10\% & two-level & Wang 1 \\
\midrule
ES 50\% & \textbf{-4.16} & -7.02 & -8.44 & -5.05 \\
ES 10\% & -4.53 & \textbf{-5.28} & -5.81 & -4.71 \\
two-level & -4.59 & -5.33 & \textbf{-5.71} & -4.73 \\
Wang 1 & -4.38 & -5.40 & -5.95 & \textbf{-4.62} \\
\bottomrule
\end{tabular}
}
\end{tabular}
\end{table}

Each hedge is best under its own preference, in twelve pairwise comparisons out of twelve in every market and by
at least $0.05$. The largest margins come from the hedge trained at $50\%$ scored deeper in the tail. Under expected shortfall at
$10\%$ the hedge trained at $50\%$ scores $1.3$ to $2.2$ lower, and under the two-level mixture $1.5$ to
$3.7$ lower. The better entropic hedge, at $\lambda = 50$, scores $0.43$ to $0.54$ below the shortfall
hedge at $10\%$.

The shortfall curves of panels b and c show where each hedge protects. Each curve is highest where its
weight charges: the hedge trained at $50\%$ near one half, the one trained at $10\%$ near one tenth, and the
two-level hedge in the deepest percent. The curves cross, so no hedge is best at every depth. The law
itself moves accordingly (panel a). The hedge trained at $50\%$ keeps more of the profit and loss near zero
and pays with a longer jump tail, while the entropic hedge joins the two-level hedge in the deepest
percent on Merton.

The two-level mixture has a reference on two markets. On Brownian motion the reference $-5.52$ passes the
check, and on Merton $-8.82$ is the value of a grid policy. At $512$ steps policy search is $2.0$ and
$1.5\%$ from them, policy search with thresholds $2.0$ and $1.4\%$, the policy gradient $1.9$ and $1.3\%$,
and the actor--critic $2.2$ and $2.0\%$; at $2{,}048$ steps policy search reaches $1.3$ and $0.9\%$. The
weight thus decides which depths of the tail a hedge protects, at the expense of the others.

\subsection{Thirty Dates and Whole Lots}
\label{subsec:horizon}

Do the learned hedges hold at thirty dates, the daily rebalancing of deep hedging
\citep{buehler2019-deep-hedging}, and under frictions without a usable derivative in the trade?
Tables~\ref{tab:T30} and~\ref{tab:lots} answer the two questions in turn.

\begin{table}[H]
\centering
\caption{Policy search at thirty dates, cost $1\%$, $2{,}048$ steps: median score and gap (\%) where a
reference exists; on Heston only the entropic utility has one.}
\label{tab:T30}
\setlength{\tabcolsep}{4pt}
\begin{tabular}{lrrrrrr}
\toprule
 & ES 50\% & ES 10\% & two-level & Wang 1 & entropic 25 & entropic 50 \\
\midrule
Brownian motion, policy search & -4.18 (0.7\%) & -4.99 (2.4\%) & -5.35 (2.8\%) & -4.59 & -3.60 (2.3\%) & -3.85 (1.7\%) \\
Brownian motion, delta hedge & -5.74 & -6.70 & -7.22 & -5.97 & -5.13 & -5.24 \\
Heston, policy search & -4.13 & -5.03 & -5.46 & -4.56 & -3.49 (2.4\%) & -3.76 (1.5\%) \\
Heston, delta hedge & -5.72 & -6.82 & -7.45 & -6.01 & -5.12 & -5.24 \\
\bottomrule
\end{tabular}

\end{table}

\begin{table}[H]
\centering
\caption{Whole lots of $\ell = 0.1$, cost $0.2\%$ and a fee $f$ per ticket on Brownian motion at thirty
dates, $2{,}048$ steps: median scores. The reference is the value of the grid policy on the lots.}
\label{tab:lots}
\begin{tabular}{lrrrr}
\toprule
 & ES 50\% & ES 10\% & entropic 25 & entropic 50 \\
\midrule
reference & $-3.25$ & $-3.83$ & $-2.77$ & $-2.89$ \\
policy search, straight-through & $-3.57$ & $-4.13$ & $-3.09$ & $-3.18$ \\
policy gradient & $-3.42$ & $-4.07$ & $-2.87$ & $-2.98$ \\
actor--critic & $-3.41$ & $-4.00$ & $-2.87$ & $-2.99$ \\
delta hedge, rounded to the lot & $-4.02$ & $-4.68$ & $-3.62$ & $-3.67$ \\
\bottomrule
\end{tabular}

\end{table}

At thirty dates policy search is $0.7$ to $2.8\%$ from the reference in the seven cells that have one,
among them $2.4$ and $1.5\%$ under the entropic utility on Heston. It is above the delta hedge on every
seed in all twelve cells, its medians by $1.38$ to $1.98$, and closes $93$ to $98\%$ of the delta hedge's
gap, with seeds within $0.022$. The gaps are larger than at ten dates with the same budget: on Brownian
motion $2.4$ against $1.1\%$ under expected shortfall at $10\%$, and $1.7$ against $0.7\%$ at
$\lambda = 50$. Each spectral preference still selects its own hedge, twelve pairwise comparisons out of twelve on both
markets. The comparison of Section~\ref{subsec:distribution} thus carries over to daily rebalancing. Trading more often costs the delta hedge most. From ten to thirty dates on Brownian motion the
reference rises from $-4.20$ to $-4.16$ at $50\%$ and from $-5.11$ to $-4.88$ at $10\%$, and policy search
from $-4.20$ to $-4.18$ and from $-5.17$ to $-4.99$, while the delta hedge, which ignores the cost, falls
from $-4.96$ to $-5.74$ and from $-5.98$ to $-6.70$.

The optimal value cannot fall when dates are added, since a thirty-date hedge may trade at every third
date only and so reproduce any ten-date hedge; the rise of the reference is consistent with the value of the additional choices. The delta hedge
has no such choice. It rebalances at every date whatever the cost, and pays the proportional cost on
three times as many trades. The larger gaps of policy search at thirty dates are consistent with its
budget: with three times as many decisions and the same number of steps, each trade receives a smaller
share of the training signal.

Whole lots and a fee reverse the order of the methods. All three beat the rounded delta hedge, itself
$22$ to $31\%$ from the reference, on every seed, with seeds within $0.05$. A fee per ticket rewards
fewer and larger trades, which the rounded delta hedge does not make, since it trades whenever the
rounded delta changes. Policy search, whose
straight-through gradient is not the gradient of the objective, is $7.7$ to $11.4\%$ from the reference
and closes $58$ to $66\%$ of the rounded delta hedge's gap. The policy gradient is $3.2$ to $6.1\%$ from it
and closes $73$ to $88\%$, and the actor--critic is $3.3$ to $5.1\%$ and closes $79$ to $89\%$. Both score
above policy search on every seed in every cell, the reverse of Table~\ref{tab:converged}, where policy
search leads the policy gradient in every cell that has both. With lots the target is one of thirteen
multiples of $\ell$, and with a fee the profit and loss is piecewise constant in the trade, so the
derivative that policy search follows carries little information. The likelihood-ratio methods never differentiate the
reward and need only its value on each path, so the jumps that defeat the straight-through gradient only
keep their exploration above a floor.

The critic helps where the signal is sparse. The actor--critic beats the policy gradient on every seed
only under expected shortfall at $10\%$ ($4.5$ against $6.1\%$), where the same-path difference carries
signal only on the paths that reach the tail; elsewhere their medians differ by less than the
actor--critic's seed spread. At $\kappa = 1$, where the critic is only a baseline, its gap widens to
$5.8\%$ there and from $3.3$ to $4.1\%$ at $\lambda = 25$, near the policy gradient's $6.1$ and $3.6\%$; at
$\kappa = 0.5$ it is $4.8$ and $3.2\%$. These two cells suggest, without establishing it, that the gain is
the credit the critic assigns across the thirty dates, which is the gain Section~\ref{subsec:ac-advantage}
anticipates. Without a usable gradient in the trade, the likelihood-ratio methods therefore lead, and the
critic adds most where the signal reaches few paths.

\subsection{A Single Distributional Critic}
\label{subsec:onecritic}

\begin{figure}[H]
\centering
\includegraphics[width=\textwidth]{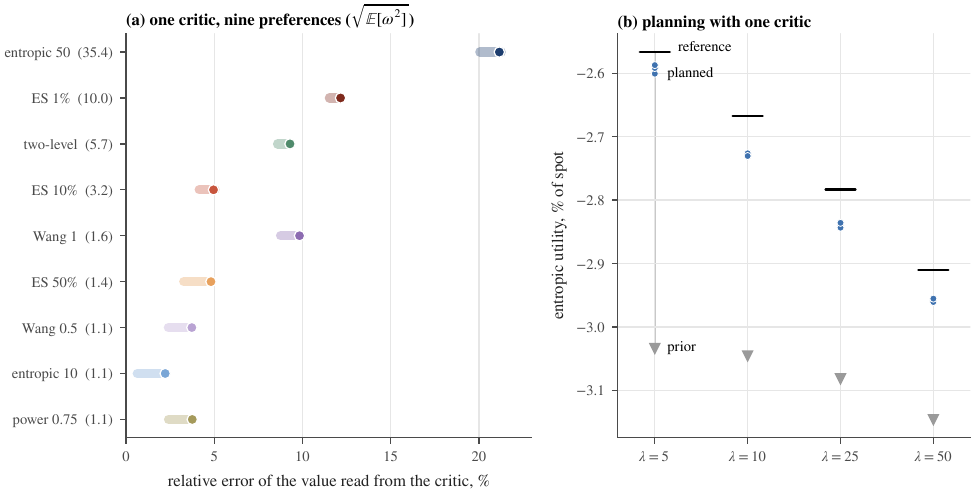}
\caption{(a) Relative error of the value read from one critic under nine preferences, sorted by
$\sqrt{\mathbb{E}[\omega^2]}$ (in parentheses), for the grid hedge of the entropic utility at
$\lambda = 10$, Brownian motion, ten dates, cost $1\%$: range over three critics and median. (b) Planning
with one critic on the lots problem at ten dates under the entropic utility.}
\label{fig:critic}
\end{figure}

Can one critic, fitted with no preference, serve preferences chosen afterwards, both when its values are
read and when a planner acts on them? Figure~\ref{fig:critic} and Table~\ref{tab:planning} test the two
uses in turn.

\begin{table}[H]
\centering
\caption{Planning with one critic under the entropic utility, lots problem at ten dates: the prior (the
delta hedge rounded to the lot), the planned hedge (median over critics) and the reference, in units of
$1\%$ of the spot, on the first $20{,}000$ test paths, and the share of the gap between prior and reference
closed.}
\label{tab:planning}
\begin{tabular}{rrrrr}
\toprule
$\lambda$ & prior & planned & reference & gap closed \\
\midrule
5 & -3.03 & -2.59 & -2.57 & 95\% \\
10 & -3.05 & -2.73 & -2.67 & 84\% \\
25 & -3.08 & -2.84 & -2.78 & 81\% \\
50 & -3.15 & -2.96 & -2.91 & 79\% \\
\bottomrule
\end{tabular}

\end{table}

To read values, a critic with $M = 64$ levels is fitted, with no preference in its loss, to the returns of
the entropic grid hedge at $\lambda = 10$. Nine preferences are applied to its outputs at the initial
state and compared with their values on the test paths (panel a). The errors grow with
$\sqrt{\mathbb{E}[\omega^2]}$, at rank correlation $0.90$, from $11.5$ to $12.2\%$ under expected shortfall at
$1\%$ to $20.1$ to $21.3\%$ under the entropic utility at $\lambda = 50$. The order is not exact, since the
Wang distortion at $a = 1$ errs by $8.8$ to $9.9\%$, but near the mean the errors fall to $0.6$ to
$3.8\%$. All nine values are optimistic.

Planning asks less of a critic than reading a value (panel b). Reading needs the critic to be right at one
state, planning only that it rank the candidates at each state, and under the entropic utility a
constant error common to all of them cancels. With the critic of
the rounded delta hedge, the planner closes $79$ to $95\%$ of the gap under the entropic utility at four
risk aversions. It is above the prior on every seed, and the three critics agree within $0.014$. It
trades $0.5$ to $2.4$ times per path, against $7.0$ for the prior and $0.5$ to $2.6$ for the reference, so it
recovers the reference's restraint and most of its score. Under expected shortfall at $10\%$, with
thresholds moved after each of three passes, it closes $86\%$ of the gap, from $-4.35$ to $-4.07$ against
$-4.02$. At $50\%$ it closes $28\%$, and $3\%$ after eight passes. There the thresholds have moved from the
prior's quantiles, where Proposition~\ref{prop:leaf} gives no ascent, and the planned hedge barely improves. One critic therefore serves preferences chosen
later, with an error that grows with the weight on the tail.

\subsection{Sample Cost of a Preference}
\label{subsec:cost}

\begin{figure}[H]
\centering
\includegraphics[width=\textwidth]{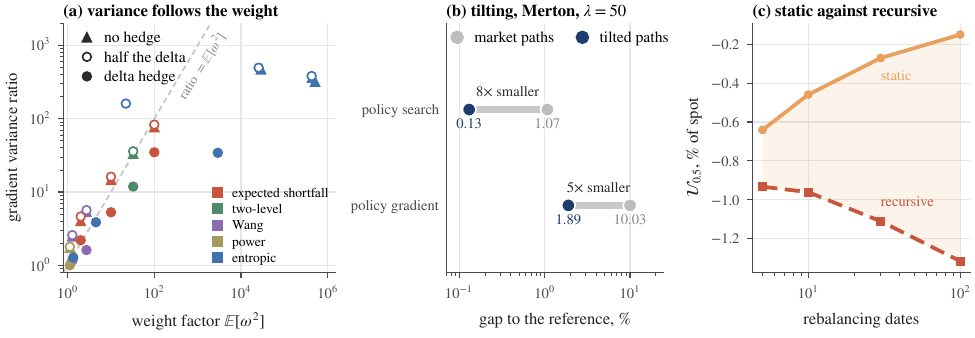}
\caption{(a) Gradient variance relative to the unweighted gradient's, for ten preferences, coloured by
family, and three fixed hedges, by marker; the dashed line is $\mathbb{E}[\omega^2]$ itself. (b) Gap to the
reference, Merton, $\lambda = 50$, $2{,}048$ steps, median of seeds $0$ to $2$, on market and on tilted
paths. (c) Expected shortfall at $50\%$, static and recursive, of the delta hedge on Brownian motion,
no costs.}
\label{fig:cost}

\vspace{4pt}
\includegraphics[width=\textwidth]{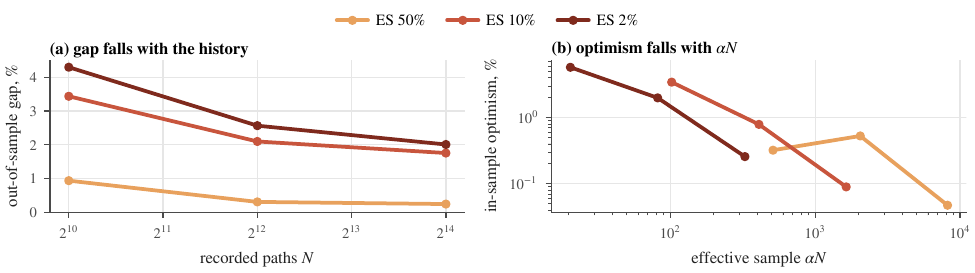}
\caption{Policy search trained on $N$ recorded paths, $1{,}024$ per step for $2{,}048$ steps, Brownian motion,
ten dates, cost $1\%$. (a)
Out-of-sample gap to the reference. (b) Excess of the in-sample over the out-of-sample score against
$\alpha N$; at $50\%$ with the largest $N$ it is $0.01\%$, drawn at $0.05\%$.}
\label{fig:history}
\end{figure}

What does a preference cost in sample paths, and can the cost be lowered? Section~\ref{subsec:estimability}
predicts that the second moment of the weight sets the cost; Figures~\ref{fig:cost} and~\ref{fig:history}
test that prediction.

The variance of the weighted gradient follows the factor (Figure~\ref{fig:cost}a). For no hedge, half the
delta and the delta on Brownian motion at thirty dates, the variance relative to the unweighted gradient
is ordered by $\mathbb{E}[\omega^2]$, at rank correlation $0.95$ to $0.98$, though not bounded by it, since
Proposition~\ref{prop:ess} bounds the variance relative to the unweighted bound. The spectral factor is the
same on every hedge, but the entropic factor depends on the hedge. At $\lambda$ equal to the inverse
standard deviation of the delta hedge's profit and loss it is $4.5$ on the delta hedge, $2.6 \cdot 10^4$ on
half of it and $4.4 \cdot 10^5$ with none.

Tilting the sampling measure lowers the cost (panel b). On Merton at $\lambda = 50$, with the jump
intensity raised fivefold and the paths reweighted by the likelihood ratio, policy search moves from
$1.07$ to $0.13\%$ from the reference, within its standard error, and the policy gradient, on every seed, from $10.0$ to $1.89\%$. Panel c
returns to the comparison of Section~\ref{subsec:time-consistency}, with the delta hedge's error in place
of the fixed variance. As the dates grow from $5$ to $100$ the static value rises from $-0.64$ to $-0.15$,
while the recursive one worsens from $-0.93$ to $-1.32$.

A finite history adds its own cost (Figure~\ref{fig:history}). On $N$ recorded paths the out-of-sample gap
falls from $0.9$ to $0.2\%$, $3.4$ to $1.8\%$ and $4.3$ to $2.0\%$ at $50\%$, $10\%$ and $2\%$ as $N$ grows
from $1{,}024$ to $16{,}384$. The effective sample is $\alpha N$, and against it the optimism at $2\%$ and
$10\%$ falls from $5.8\%$ to $0.09\%$, while at $50\%$ it stays below $0.6\%$. The second moment of the
weight thus predicts the cost of a preference, and a tilted sampling measure reduces it.

\section{Related Work}
\label{sec:related-work}

Risk-sensitive hedging with learned policies sits between deep hedging, distributional reinforcement
learning and risk-sensitive dynamic programming. In all of them the preference is fixed before training,
which is where this paper departs from each (Table~\ref{tab:related}); one paragraph collects the
estimation results the paper relies on.

\begin{table}[H]
\centering
\caption{Lines of related work. Preference fixed: chosen before training. Static: the preference is
applied once to the terminal profit and loss; nested: applied date by date; per state: read from a
critic at each state.}
\label{tab:related}
\begin{tabularx}{\textwidth}{@{}P{2.7cm}LP{1.65cm}P{2.55cm}P{1.95cm}l@{}}
\toprule
line of work & objective & preference fixed & learned object & static or nested & references \\
\midrule
deep hedging, static objective & optimised certainty equivalents, penalties & yes & policy, or policy and value & static & \citep{buehler2019-deep-hedging,carbonneau2020-deep-hedging-long-term-derivatives,imaki2021-no-transaction-band-network,peng2024-risk-sensitive-contract-unified-rl-option-hedging} \\
actor--critic hedging & entropic, monetary utilities & yes, within a family & policy and value & nested; static if entropic & \citep{murray2022-deep-hedging-continuous-rl,buehler2022-deep-bellman-hedging,marzban2023-deep-rl-expectile-risk} \\
distributional critics in hedging & shortfall, value at risk, mean--deviation & yes & quantiles & per state & \citep{cao2023-gamma-vega-hedging,sharma2024-distributional-rl-autocallable,zhang2025-tail-safe-hedging} \\
distributional reinforcement learning & expectation, distortions & yes, per agent & return distribution & per state & \citep{bellemare-2017-c51-distributional-rl,dabney2018-quantile-regression,dabney2018-implicit-quantile-networks} \\
static risk in decision processes & shortfall, spectral, certainty equivalents & yes & value on the enlarged state & static & \citep{bauerle2011-mdp-average-value-at-risk,bauerle2021-spectral-risk-measures-mdp,wang2025-oce-risk-sensitive-reductions} \\
nested risk & Markov and dynamic convex measures & yes & value & nested & \citep{ruszczynski2010-risk-averse-dynamic-programming,coache2024-rl-dynamic-convex-risk} \\
\midrule
this paper & coherent spectral, entropic & critic: no; policy: yes & quantiles, no preference in the loss & static & Sections~\ref{sec:formulation}--\ref{sec:experiments} \\
\bottomrule
\end{tabularx}
\end{table}

Deep hedging trains a network policy against a convex risk measure of the terminal profit and loss
under trading costs \citep{buehler2019-deep-hedging}. The measure is an optimised certainty equivalent,
the entropic utility or expected shortfall, illustrated on a call hedged daily over thirty days in a
Heston market. Later work varies the instrument, the market or the architecture:
quadratic and semi-quadratic penalties for long-dated guarantees \citep{carbonneau2020-deep-hedging-long-term-derivatives}, a
no-transaction band built into the network, which clamps the previous hedge into a band it outputs
\citep{imaki2021-no-transaction-band-network}, and a
mean-plus-deviation objective with two moment critics \citep{cao2021-deep-hedging-reinforcement-learning}.
The objective itself has been changed to equal risk pricing \citep{carbonneau2021-equal-risk-pricing-deep-hedging},
recursive expectiles \citep{marzban2023-deep-rl-expectile-risk} and the static shortfall of the terminal
profit and loss \citep{peng2024-risk-sensitive-contract-unified-rl-option-hedging}.
Each of these fixes the preference in every learned object. Here the critic carries none, and only the
policy is trained per preference.

Actor--critic methods have entered hedging mainly through monetary utilities.
\citet{murray2022-deep-hedging-continuous-rl} fit a critic to the entropic certainty equivalent and pass
the risk aversion, drawn for each episode, to actor and critic, so one network covers a range of risk
aversions. \citet{buehler2022-deep-bellman-hedging} derive a Bellman equation for monetary utilities,
which equals the static objective only for the entropic utility and the expectation. Their model is
trained once and hedges any initial portfolio, a generality along the portfolio rather than the
preference.
In both, the critic depends on the preference. The critic of this paper carries none, so a preference
outside the entropic family needs no new critic.

Distributional critics have also been used in hedging. Quantile critics have hedged the gamma and vega of an option book \citep{cao2023-gamma-vega-hedging} and a
portfolio short an autocallable note \citep{sharma2024-distributional-rl-autocallable}, each under a fixed
objective. In the second, one value-at-risk level alone drives the agent to the delta--gamma hedge: the
preference selects the hedge. \citet{zhang2025-tail-safe-hedging} pair an implicit quantile critic with an update weighted by
conditional value at risk and a constraint layer. In quantum deep hedging
\citep{cherrat2023-quantum-deep-hedging} the distribution of future paths is carried in the amplitudes of a
quantum state and the exponential utility is read from an observable.
Where these works read a risk measure from the critic, they read it state by state, and that does not
optimise the static objective (Section~\ref{subsec:time-consistency}).

In distributional reinforcement learning, the law of the return is represented on a fixed support \citep{bellemare-2017-c51-distributional-rl},
whose sample-based algorithm converges \citep{rowland2018-analysis-categorical-distributional-rl}, by
quantiles at fixed levels \citep{dabney2018-quantile-regression}, by a network evaluated at sampled levels
\citep{dabney2018-implicit-quantile-networks}, or by learned levels
\citep{yang2019-fully-parameterized-quantile-function}; see \citet{bellemare2023-distributional-rl}. The
implicit quantile network applies a distortion, among them the Wang, power and shortfall distortions, by
resampling the levels of one network rather than through its loss.
The resampling construction is that of Dabney et al. Section~\ref{sec:formulation} collects, with their sources, the
conditions under which it is well posed, adds the entropic gradient, and measures the cost of a
preference by the second moment of its weight.

Reading a preference state by state has a known flaw. Applying a distortion at every state converges to neither the static nor the nested expected shortfall
\citep{lim2022-distributional-rl-risk-sensitive-policies}. The general mechanism is a functional not
indifferent to mixtures \citep{avilapires2025-distributional-dynamic-programming}. Augmenting the state by
a risk level \citep{chow2015-cvar-risk-sensitive-robust-decision-making} is suboptimal, while augmenting it
by the accumulated reward is unaffected \citep{hau2023-dp-decompositions-static-risk}; a decomposition holds
for the quantile itself \citep{hau2024-q-learning-quantile-mdps}.
Section~\ref{subsec:time-consistency} states the failure for hedging.

Augmenting the state by the accumulated reward is the established route to a static objective in a decision process
\citep{bellemare2023-distributional-rl}. \citet{bauerle2011-mdp-average-value-at-risk} treat the average
value at risk and \citet{bauerle2021-spectral-risk-measures-mdp} spectral measures, solving an inner
decision process on the extended state and, numerically, an outer problem over a threshold function. The outer problem adds to the cost of the
route: for a continuous spectrum it runs over a function, and for a mixture over one threshold per level. Policy-gradient and actor--critic
methods on such a state \citep{chow2014-cvar-optimization-in-mdps}, regret bounds
\citep{bastani2022-regret-risk-sensitive,chen2024-provable-risk-sensitive-distributional-rl}, a reduction of
optimised certainty equivalents to risk-neutral learning \citep{wang2025-oce-risk-sensitive-reductions}
and alternating schemes \citep{moghimi2025-beyond-cvar-spectral-risk,moghimi2025-risk-sensitive-actor-critic-srm}
follow the same route. Policy gradients exist for shortfall \citep{tamar2014-cvar-sampling}, coherent
measures \citep{tamar2015-policy-gradient-coherent-risk,xiao2024-distributional-policy-gradient-coherent-risk}
and distortion functionals under model uncertainty \citep{jaimungal2021-robust-risk-aware-rl}.
Proposition~\ref{prop:es-dp} is an instance of this route. The frozen problem of
Proposition~\ref{prop:minorant} is its inner problem at the current policy, used so that any risk-neutral
method applies without solving the inner problem first.

The nested alternative has a longer history. Risk-sensitive decision processes under an exponential utility go back to \citet{howard1972-risk-sensitive-mdp},
with regret analysed through the exponential Bellman equation \citep{fei2021-exponential-bellman-equation}.
Time-consistent dynamic risk measures are nested \citep{ruszczynski2010-risk-averse-dynamic-programming};
among law-invariant, relevant, cash-invariant ones only the entropic family is time consistent
\citep{kupper2009-law-invariant-time-consistent}. Policies
have been learned for dynamic convex measures \citep{coache2024-rl-dynamic-convex-risk},
conditionally elicitable spectral ones \citep{coache2022-conditionally-elicitable-dynamic-risk} and measures
defined on distributions \citep{cheng2024-risk-averse-mdp-distributional-lens}. Nesting
lets dynamic programming apply directly, at the price of changing the objective.
Outside the entropic case a nested objective differs from the static one, and its value depends on the
rebalancing frequency (Section~\ref{subsec:time-consistency}); this paper keeps the static objective.

Elicitability results support the order of estimation. Among spectral functionals only the mean is elicitable
\citep{ziegel2016-coherence-and-elicitability}, while quantiles are. Quantile regression with networks
attains minimax rates up to logarithmic factors
\citep{padilla2020quantile-regression-deep-relu-minimax}; monotonicity in the level can be restored by
rearrangement \citep{chernozhukov2010quantile-probability-curves-without-crossing}; expected shortfall is
jointly elicitable with value at risk \citep{fissler2016-higher-order-elicitability-osband}.

Classical shortfall hedging is related in name more than in substance. Quantile hedging
\citep{follmer1999-quantile-hedging} maximises, under a capital constraint, the probability that the hedge
succeeds, and efficient hedging \citep{follmer1999-efficient-hedging-shortfall-risk} minimises the expected
value of a loss function of the shortfall. With a linear loss this is the expected amount of the
shortfall, which shares a name with expected shortfall but is a different functional, and networks
replicate the resulting claims \citep{hou2021-deep-partial-hedging}. Neither criterion is a spectral
functional.
Under proportional costs \citet{leland1985-option-pricing-replication-transactions-costs} adjusts the
Black--Scholes strategy, \citet{boyle1992-option-replication-discrete-time} replicate in a binomial model,
and the exponential-utility hedge holds a no-transaction band whose width scales as the cube root of the
cost \citep{whalley1997-asymptotic-hedging-transaction-costs}. \citet{itkin2026-when-hedging-changes-payoff} replicate under price impact, which (E1) excludes here. All
the preferences give a replicated book the same value; this paper treats the risk left when replication fails.
Tree
search has been applied to hedging for an expected utility
\citep{szehr2021-mcts-hedging,maggiolo2025-deep-hedging-non-convexity-alphazero}; the planner of
Section~\ref{subsec:planning} instead searches one date ahead with the preference at its leaves.

In summary, the paper takes its interface from distributional reinforcement learning and its objective from the
static risk literature. It states what a weight must satisfy for the two to fit, and it organises the routes to the static objective around one device, the expected-utility minorant. The
objective is differentiated as a whole, or solved on the enlarged state, or, for the entropic utility, on
the market-and-inventory state. In every case the critic carries no preference.

Table~\ref{tab:closest} relates each algorithm of Section~\ref{sec:algorithms} to its closest prior method. The
last column states what each algorithm adds; everything else in it is taken from the cited work, and the
credit for it stays there.

\begin{table}[H]
\centering
\caption{Each algorithm of Section~\ref{sec:algorithms} against its closest prior method.}
\label{tab:closest}
\begin{tabularx}{\textwidth}{@{}P{3.8cm}P{6.2cm}L@{}}
\toprule
algorithm & closest prior method & difference \\
\midrule
policy search (\ref{subsec:policy-search}) & deep hedging \citep{buehler2019-deep-hedging} & rank weights serve every spectral preference \\
policy gradient (\ref{subsec:policy-gradient}) & shortfall and coherent policy gradients \citep{tamar2014-cvar-sampling,tamar2015-policy-gradient-coherent-risk} & a baseline rolled on the same market path \\
actor--critic (\ref{subsec:ac-advantage}) & entropic actor--critic \citep{murray2022-deep-hedging-continuous-rl} & a distributional critic with no preference \\
shortfall thresholds (\ref{subsec:shortfall}) & joint gradient in policy and threshold \citep{buehler2019-deep-hedging} & the exact threshold step, then a frozen block \\
shortfall programme (\ref{subsec:es-dp}) & extended-state recursion \citep{bauerle2021-spectral-risk-measures-mdp} & solved on the cushion, one pass for every threshold \\
entropic programme (\ref{subsec:entropic-dp}) & risk-sensitive recursion \citep{howard1972-risk-sensitive-mdp} & solved on a grid with costs and inventory \\
planning (\ref{subsec:planning}) & tree search \citep{szehr2021-mcts-hedging} & depth one, distributional leaves, preference at trading time \\
\bottomrule
\end{tabularx}
\end{table}

\section{Discussion and Conclusion}
\label{sec:discussion}

Table~\ref{tab:glance} shows what changes when the preference is applied after training.

The first lesson is that the preference is not a detail of training. It changes the optimal hedge, its
trading frequency and, under expected shortfall, its dependence on the past (Section~\ref{subsec:exact}),
and a hedge trained for one preference is measurably worse under another
(Section~\ref{subsec:distribution}). The static objective also has to be solved as such. Applying expected
shortfall recursively loses $8.4\%$ against the static reference, and its value drifts with the number of
dates (Sections~\ref{subsec:exact} and~\ref{subsec:cost}).

The second lesson is that the method should match the simulator. With a differentiable reward, policy
search reaches the references within $2\%$ (Section~\ref{subsec:learned}). Without one, as with whole lots
and a fee, the likelihood-ratio methods lead, and the critic adds credit across dates
(Section~\ref{subsec:horizon}). One critic is then enough to choose the preference late. Values read from
it are accurate near the mean and degrade with the weight on the tail, and a planner acting on it closes
most of the gap under the entropic utility and under expected shortfall at $10\%$
(Section~\ref{subsec:onecritic}). Finally, the cost of a preference is predictable: the second moment of
the weight orders both the gradient variance and the critic's error, and a tilted sampling measure lowers
it (Section~\ref{subsec:cost}).

For a desk, the practical consequence is a division of labour. A critic is fitted once to the book's
hedge, with no preference, and the preference is chosen afterwards, either read from the critic, with an
error that grows toward the tail, or applied by the planner when the trade is made. Each preference
still needs its own trained policy, and for a spectral one the cost of training it is known from
$\lVert w\rVert_2^2$ before any path is simulated.

These conclusions rest on a deliberately narrow test. The experiments hedge one option, at a thirty-day
maturity, with ten or thirty dates and a small network, on simulated paths only; no market data are used.
The policies are trained per preference, so only the critic, and through it the planner, carries none.
Each configuration has three seeds, and differences between methods below the seed spread are not
resolved; neither are gaps of the order of the standard error of a reference, $0.1$ to $0.3\%$ of its value on
Brownian motion and $1.3\%$ on Merton at $\lambda = 50$.
Among time-consistent alternatives only the recursive shortfall hedge is compared with the static
objective. Nor are the planner and the trained policies compared on one problem.

The references carry their own limits. They are self-consistent to within $1\%$ only on Brownian motion
at ten dates with a cost of $1\%$, except at $\lambda = 25$, and on Merton at $\lambda = 25$. On Heston under expected shortfall, and on
Merton under the two-level mixture and at $\lambda = 50$, they fail that check, and the Wang distortion has no reference at all. The entropic values are those of the
simulated laws (Section~\ref{subsec:entropic}); under Merton at $\lambda = 50$ they depend on how far into the
jump tail that law reaches, which is also where the grid and the test paths disagree.
Each method also has a weak point that the experiments expose. Policy search needs a differentiable
reward, and with whole lots and a fee the likelihood-ratio methods score above it. Values read from a
critic are least accurate under the preferences that weight the tail most, and all nine values read were
optimistic. Planning under expected shortfall at $50\%$ closes $28\%$ of the gap, and $3\%$ after eight
passes, once the thresholds have moved from the prior's quantiles, where Proposition~\ref{prop:leaf}
gives no ascent.

Several problems remain open. The first is a computable programme for a continuous spectrum, which
would give the Wang distortion a reference; for such a spectrum the outer optimisation of
\citet{bauerle2021-spectral-risk-measures-mdp} runs over a function rather than finitely many thresholds.
The second is whether planning, repeated with a critic refitted to each improved hedge, converges when
the critic is approximate, since Proposition~\ref{prop:leaf} guarantees one improvement only with exact
leaves. The third is a sampling measure chosen from the preference itself, which would make the cost of a
preference a quantity the algorithm controls. Two further problems concern scale. Books of several claims and several instruments put the grid
references out of reach, so the self-consistency check would need a substitute, and market data would
replace the simulator of (E1) by a fitted model or by recorded paths, whose finite number
Section~\ref{subsec:cost} shows to matter. Claims with discontinuous or path-dependent payoffs, such as a digital or a barrier option, test the
continuity assumptions behind Proposition~\ref{prop:policy-gradient} and the rank weights and, for a
barrier, add a coordinate to the grid, the programmes of Section~\ref{sec:algorithms} otherwise carrying
over unchanged. They are the natural next experiment. The last problem is an actor driven by the critic's
action-gradient.
Two methods that act on the critic were tried and dropped for failing. A deterministic actor--critic closed $52$ to $67\%$ of
the gap under the entropic utility on two markets and failed at low cost, with whole lots and on Merton,
and distributional Q-learning failed on Merton. A critic accurate in its slope, and not only in its
values, is the missing piece.

To conclude, the coherent preferences of this form are the mixtures of expected shortfalls. Those with
bounded weight can be read from a learned law, and the entropic utility is reached through its gradient.
A lower bound that touches each of them at the current policy lets a risk-neutral method, applied to the
frozen problem, improve the preference, and it organises the seven algorithms. The experiments test the
steps a grid can check: the references are solved, the learned hedges approach them, and the planner acts
on a critic fitted before the preference was chosen.

\section*{Acknowledgements}

The work was carried out with an AI research assistant, Claude \citep{anthropic2026-claude}, working in the
authors' repository, where it read and edited files and ran code. The authors posed the questions, chose between the alternatives the model proposed, and
decided what entered the paper. The model assisted at every stage: it built a structured map of the
deep-hedging and distributional reinforcement-learning literatures, proposed formulations and proofs,
wrote the simulation, training and dynamic-programming code, ran the experiments, and drafted the text,
which the authors revised.

The results rest on checks rather than on the model's output. Statements quoted from the literature were
matched by machine against their sources, and the hypotheses of the cited results were checked against
the setting of the paper. Experiments were specified in advance by written protocols with predictions and
refutation thresholds, every table and figure was generated by script from recorded results, and the
references of Section~\ref{sec:experiments} were checked for self-consistency, with the outcome reported.
Proofs, numbers and citations were reviewed by the authors and in separate review passes run with the
model, and claims that failed review were withdrawn. The model's typical errors were overstated claims,
generalisations from small cases, misattributed results, and code returning plausible but wrong numbers;
systematic checks caught these more reliably than reading did.

This account records how the results were produced and checked. It is not a study of AI-assisted research
and does not separate the model's contribution from what the authors would have done alone. The authors take full
responsibility for the content of the paper.

\samepagesection
\bibliographystyle{unsrtnat}
\bibliography{references}

\end{document}